\documentclass[acmlarge]{acmart}
\AtBeginDocument{%
  }

\setcopyright{cc}
\setcctype{by}
\acmJournal{IMWUT}
\acmYear{2026} \acmVolume{10} \acmNumber{3} \acmArticle{166}
\acmMonth{9} \acmDOI{10.1145/3831976}

\usepackage{graphicx}
\usepackage{subcaption}
\usepackage{amsmath}
\usepackage{pifont}
\usepackage{hyperref}

\usepackage[most]{tcolorbox}
\usepackage[normalem]{ulem}
\usepackage{xcolor}
\usepackage{mdframed}
\usepackage{xspace}
\usepackage[utf8]{inputenc}
\usepackage{enumitem}
\newcommand{\name}{Pebbl\xspace}
\usepackage{multirow}
\usepackage{multicol}
\usepackage{booktabs}   
\usepackage{array}  
\usepackage{arydshln}
\usepackage{fancyvrb}  
\usepackage{lipsum}
\usepackage{listings}
\usepackage{xcolor}
\usepackage{booktabs}
\usepackage{tabularx}
\usepackage{array}
\usepackage{ragged2e}

\newcolumntype{L}[1]{>{\RaggedRight\arraybackslash}p{#1}}
\newcolumntype{Y}{>{\RaggedRight\arraybackslash}X}

\begin{document}

\title[Towards Wearable Opportunistic Crowdsensing]{Towards Wearable Opportunistic Crowdsensing for Open-Vocabulary Activity Data Collection Through User-Scheduled Trigger-Action Routines}

\author{Zeyu Wang}
\authornote{Both authors contributed equally to this research.}
\email{wang-zy23@mails.tsinghua.edu.cn}
\orcid{0009-0007-5048-1665}
\author{Yingke Ding}
\authornotemark[1]
\email{dyk21@mails.tsinghua.edu.cn}
\orcid{0009-0006-3618-9492}
\affiliation{%
  \institution{Tsinghua University}
  \country{China}}

\author{Mingze Gao}
\affiliation{%
  \institution{Tsinghua University}
  \country{China}}
\email{gaomingze2022@gmail.com}

\author{Zhuolun Ren}
\affiliation{%
  \institution{Renmin University of China}
  \country{China}
}

\author{Alex Mariakakis}
\affiliation{%
 \institution{University of Toronto}
 \country{Canada}}

\author{Yuanchun Shi}
\orcid{0000-0003-2273-6927}
\affiliation{%
  \institution{Tsinghua University}
  \department{Beijing National Research Center for Information Science and Technology (BNRist)}
  \country{China}
}
\affiliation{%
  \institution{Qinghai University}
  \country{China}
}

\email{shiyc@tsinghua.edu.cn}

\author{Yuntao Wang}
\authornote{Corresponding author.}
\orcid{0000-0002-4249-8893}
\affiliation{%
  \institution{Tsinghua University}
  \country{China}
}
\affiliation{%
  \institution{Qinghai University}
  \country{China}
}
\email{yuntaowang@tsinghua.edu.cn}

\renewcommand{\shortauthors}{Wang et al.}

\begin{abstract}

Collecting richly labeled wearable activity data in everyday settings remains difficult because retrospective annotation is costly and often imprecise. Prior data collection apps rely on a labor-intensive self-reporting strategy and primarily treat participants as crowd labelers. We present \name, a feasibility-stage system that incentivizes in-situ labeling through opportunistic crowdsensing. \name lets users author trigger-action recipes on a smartphone and receive just-in-time reminders for beneficial actions when a trigger is detected. In the prototype, triggers are a limited set with four common audio cues, while actions are described in open-vocabulary natural language. Each confirmed execution yields a short sensor window with explicit start/end boundaries and a user-authored action label. We evaluate \name through an expert workshop with wearable Human Activity Recognition (HAR) researchers ($N=6$), a within-subject in-lab study ($N=21$), and a pilot deployment ($N=8$). Experts viewed the approach as lower burden and more ecologically valid than common labeling workflows. In the lab, \name produced reliable execution logs under controlled conditions (recall $=97.30\%$, precision $=97.15\%$) and was preferred over comparison workflows on perceived burden and confidence. The pilot deployment shows that the interaction and sensing pipeline can function in free-living use, while surfacing practical constraints such as false triggers and context dependence. Overall, \name represents a step toward a low-burden, distributable collection approach of user-contributed wearable activity data.

\end{abstract}

\ccsdesc[500]{Human-centered computing~Human computer interaction (HCI)}
\ccsdesc[500]{Human-centered computing~Ubiquitous and mobile computing}

\received{1 February 2026}
\received[revised]{1 May 2026}
\received[accepted]{6 July 2026}

\keywords{Activity Recognition, Fogg Behavior Model, Smartwatch Application, Dataset Collection, Tiny-Habit}

\maketitle

\section{Introduction}

\begin{figure}
  \includegraphics[width=\textwidth]{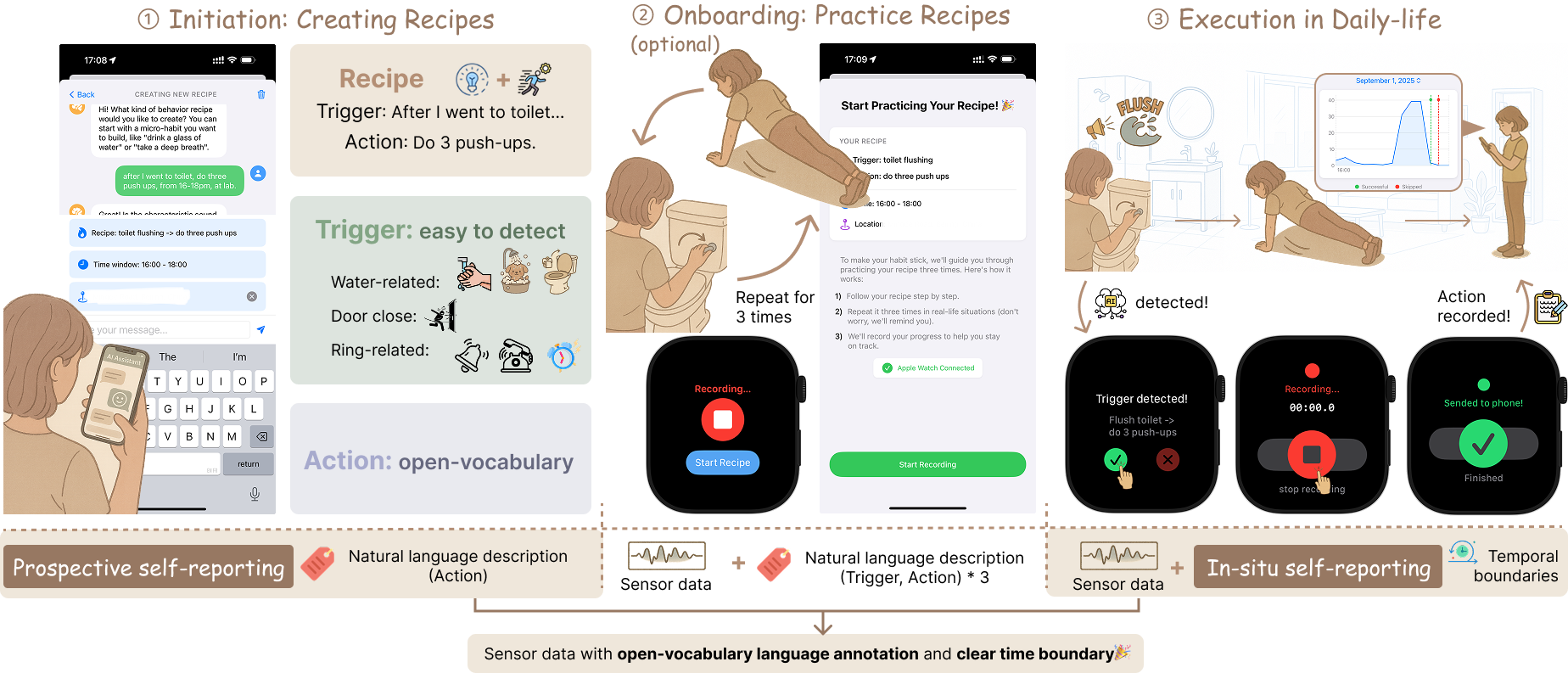}
  \Description{Composite illustration of a three-stage workflow for \name. In the first stage, the user uses a phone to chat with an AI assistant, choosing an audio trigger such as toilet flushing and a natural-language action like doing three push-ups. In the second stage, the user rehearses the recipe while the watch records practice data; in the third stage, the user receives a just-in-time watch notification when the trigger occurs in daily life, confirms or skips the action, and the watch records a short, timestamped sensor window. A strip along the bottom contrasts prospective self-reporting during recipe creation with in-situ self-reporting that yields sensor data with clear temporal boundaries.}
  \caption{\textbf{Overview of \name's interaction and data flow.} \name is an opportunistic crowdsensing system that couples data collection with reminders and trackers for beneficial actions. This figure demonstrates \name's key components. \textbf{(1) Initiation:} An Large Language Model (LLM)-powered agent helps users author recipes by specifying an audio-based \textbf{Trigger} with time/location conditions, and an \textbf{Action} described in open-vocabulary natural language (e.g., ``do three push-ups''). \textbf{(2) Onboarding:} The app guides users to rehearse the recipe on the watch and record the practice data with weak annotations. \textbf{(3) Execution in daily life:} When the trigger is detected, \name provides a just-in-time watch prompt and the user confirms whether they will complete the corresponding action. Each execution yields a timestamped, precisely bounded sensor window (e.g., audio/IMU) paired with an open-vocabulary action label, enabling in-situ self-reporting with clear temporal boundaries. }
  \label{fig:teaser}
\end{figure}

Large-scale, richly labeled datasets have driven major advances in vision and language modeling~\cite{deng2009imagenet, raffel2020exploring, schuhmann2022laion}, but wearable human activity recognition (HAR) still lacks similarly broad and well-annotated data resources~\cite{chan2024capture, yin2024systematic}. This gap is especially pronounced for fine-grained activities performed in everyday life. Controlled lab protocols yield clean labels but limited realism~\cite{alharbi2018can}, while free-living data often requires burdensome post-hoc annotation process that is time-consuming and vulnerable to recall and segmentation errors~\cite{grauman2022ego4d, grauman2024ego, wang2024dreamcatcher}. In-situ self-report can reduce recall bias by asking people to label their own activities as they happen or shortly thereafter~\cite{vaizman2018extrasensory, kim2022mymove}. However, current systems still largely frame participants as annotators performing research labor. This limits sustainability: even when labels are valuable to researchers, the interaction often remains effortful and provides little direct value to the people generating the data.

We investigate a different design direction: can wearable activity labeling be embedded in an experience that users already find useful? To explore this question, we present \name, a smartphone-smartwatch system that combines user-authored trigger-action routines with just-in-time watch prompts. Users specify a trigger, a target beneficial action; when the trigger is detected, the watch reminds the user to perform the action and records a bounded sensor window if they confirm. Available triggers are a limited set that are detectable cues, while actions are described in open-vocabulary natural language. An example interation flow is shown in Figure~\ref{fig:teaser}.

\name is inspired by trigger-action routines used in behavior-change systems~\cite{fogg2009behavior, fogg2020tiny, ur2016trigger}. We leverage this framing not to claim that \name is a validated habit-formation intervention, but to give users a practical reason to engage with in-situ labeling: the system can function as a reminder and lightweight tracker for personally meaningful micro-actions. This design naturally aligns with \textbf{wearable opportunistic crowdsensing}~\cite{campbell2006people, lane2010survey} that capitalizes on users’ everyday routines with minimal explicit action.

After implementing \name for iPhones and Apple Watches, we conducted three studies with the varied stakeholders associated with our target domain. As a data-collection approach, \name involves two central stakeholder groups: researchers who may use the resulting wearable activity data, and users who produce the labeled sensor windows through usage. First, we held a workshop with wearable HAR researchers to confirm that \name had preferable qualities relative to existing data collection approaches \textbf{(RQ1)}. We then conducted a within-subjects lab study comparing \name to both baseline annotation tools and a standard tracking app. Participants served as both \name app users and dataset annotators, allowing us to assess both data labeling reliability~\textbf{(RQ2)} and their tracking outcomes~\textbf{(RQ3)}. Finally, we conducted an in-the-wild study to understand \name's affordance in real-world settings. 

Our findings suggest that \name is preferred by HAR researchers and annotators as a data collection platform, as well as by users as an app for encouraging and tracking beneficial behaviors. Six experts endorsed \name's potential for scalability, ecological validity, and low burden, while 21 participants confirmed that they preferred \name for its low mental/physical workload and just-in-time prompts. Diagnostic logs with precise timestamps and explicit options to skip activities were well-received features, and \name yielded higher label reliability relative to all baselines. Through a pilot deployment with 8 participants, we show that \name's interaction and sensing pipeline can function well in short-term free-living use, while surfacing practical constraints around false triggers and context dependence.

Our contributions are:
\begin{itemize}
    \item \textbf{An opportunistic crowdsensing framework for wearable HAR data collection via user-scheduled trigger-action routines.} We present \name, a feasibility-stage framework that explores a path toward gathering timestamped, labeled activity data at scale while promoting beneficial actions.
    \item \textbf{A full-stack mobile–wearable system and staged empirical evaluation.} We implemented \name on an iPhone and an Apple Watch and evaluated it through expert elicitation, a controlled lab study, and a pilot deployment.
    \item \textbf{Empirical findings, design implications, and limitations for deployment.} We show that \name{} can reduce perceived burden and produce reliable execution logs in controlled settings, and derive implications and scope boundaries for broader deployment.
\end{itemize}

\section{Related Work}

\subsection{Labeling Wearable Activity Data in the Wild}

Wearable HAR datasets are commonly created through three families of labeling workflows: scripted collection in controlled settings~\cite{weiss2019wisdm, reiss2012introducing, chavarriaga2013opportunity}, free-living recording with post-hoc annotation~\cite{chan2024capture, grauman2022ego4d, xu2023towards}, and free-living recording with participant self-report~\cite{vaizman2018extrasensory, minor2025feature, kim2014space}. Controlled protocols offer clean labels but limited ecological validity. Post-hoc annotation supports naturalistic behavior but is costly and often imprecise at fine temporal scales. Self-report reduces recall delay and gives participants semantic access to their own actions, but still places ongoing annotation burden on them. Prior systems in this space have asked users to periodically enter high-level labels, receive automatic label suggestions, or verbally describe activities shortly after completion~\cite{minor2025feature, vaizman2018extrasensory, kim2022mymove}. Our work builds most directly on this third category. The key difference is that \name{} does not present labeling only as research work; instead, it embeds confirmation and logging into a user-authored trigger-action tracking flow.

\subsection{Embedding Annotation in Personally Valuable Experiences}

A longstanding Human-Computer Interaction (HCI) strategy is to collect data while giving users something valuable in return, such as engagement, utility, accessibility, or reduced effort. One thread embeds annotation into rewarding experiences: gamified systems ask users to solve puzzles, play Extra-Sensory Perception (ESP) guessing games, or compete in multi-participant scenarios while producing labels~\cite{cooper2010predicting, kim2014space, von2004labeling, von2006peekaboom, law2009input, tuite2011photocity}, while utility-oriented systems such as MovieLens, reCAPTCHA, and VizWiz exchange user contributions for recommendations, login utility, or real-time accessibility support~\cite{harper2015movielens, von2008recaptcha, bigham2010vizwiz}. A second thread is opportunistic crowdsensing, which leverages sensor-rich mobile devices to collect data with minimal user input and has been applied to smart cities, environmental monitoring, public health surveillance, and mobile health~\cite{chi2019crowdsourcing, lane2008urban, dutta2016airsense, sadilek2013crowdphysics, mankodiya2022real}.

These ideas are less developed for wearable time-series sensing, where data is harder for end users to inspect directly. \name{} adopts the same high-level principle: if a labeling interaction also helps users remember, track, or reflect on personally meaningful actions, they may be more willing to sustain it while producing useful labels.

\subsection{Trigger-Action Routines and Contextual Reminders}

Behavior-tracking and behavior-change systems often support adherence through self-tracking, reminders, or contextual prompts. Popular habit-tracking apps monitor task completion~\cite{apple-habit-search}, but prior reviews found that many focus on self-tracking and reminders while offering limited support for context-driven cues that help routines develop~\cite{stawarz2015beyond}. Event-based reminders and Just-in-Time Adaptive Interventions (JITAIs) show how prompts can be coupled to sensed or inferred contexts, including interventions for sedentary behavior~\cite{klasnja2019efficacy, cambo2017breaksense}, eating~\cite{rahman2016predicting, chen2024investigating}, smoking~\cite{saleheen2015puffmarker, stone2024presenting}, procrastination~\cite{arakawa2023catalyst}, and deep breathing~\cite{sadprasid2024leveraging}. Related in-situ sensor-triggered prompt systems include on-body camera agents~\cite{fang2025mirai}, which differ from \name{} in relying on visual context rather than user-authored trigger-action routines.

We draw specifically on the Fogg Behavior Model (FBM) and tiny-habit recipes~\cite{fogg2009behavior, fogg2020tiny}. The FBM frames behavior as occurring when motivation, ability, and a trigger coincide; tiny-habit recipes operationalize this idea by anchoring a small action to a specific existing routine, such as ``After I flush the toilet, I will do 3 push-ups.'' \name{} adapts this trigger-action structure for wearable data labeling: the trigger helps surface an in-situ prompt, while the user-authored action provides a prospective label for the recorded sensor window. We do not evaluate \name{} as a habit-formation intervention or propose a new behavior-change theory; rather, we use this framing to create a low-burden path to activity labeling that is also personally meaningful to users.

\section{\name Design and Implementation}

\begin{figure}[ht]
    \centering
    \includegraphics[width=0.9\linewidth]{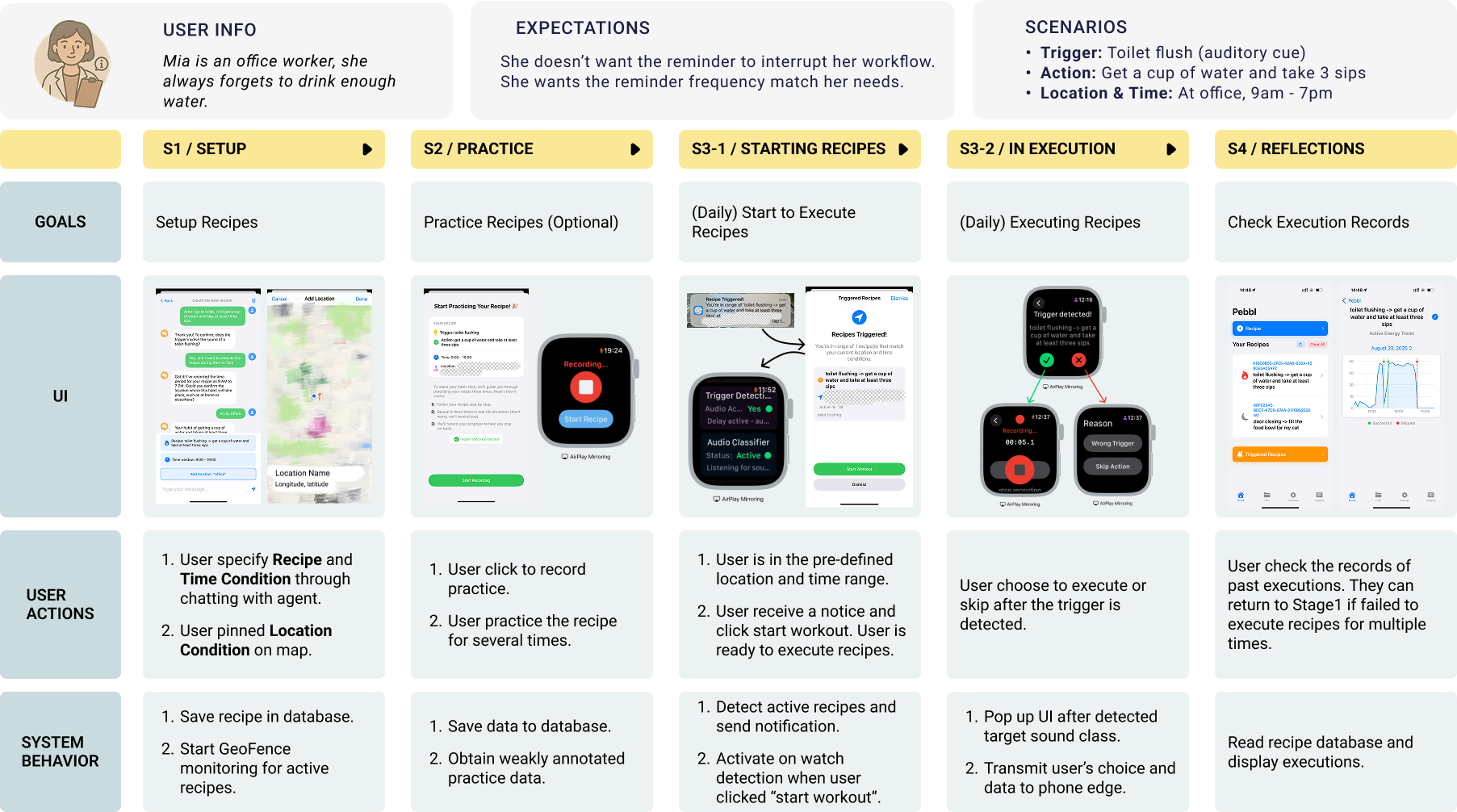}
    \caption{An experience map illustrating a typical user's journey with \name.}
    \label{fig:methods_user_journey}
\end{figure}

\subsection{Design Overview}

\name is designed around four stages: recipe creation, optional practice, in-situ execution, and reflection, as summarized in \autoref{fig:methods_user_journey}. During recipe creation, the user specifies (1) a trigger, (2) a target action, and (3) optional time and location constraints. In the current prototype, triggers must correspond to one of four non-speech audio classes: door closing, toilet flushing, running water, or ringtone/alarm. Actions are authored in free-form natural language and are intended to be short, feasible, and immediately executable after the trigger.

When a recipe becomes active, \name monitors for the trigger within the specified context. If the trigger is detected, the watch presents a lightweight prompt asking the user whether they will perform the action. If they confirm, recording starts; if they skip, the skip is logged. The user taps again when the action is complete, producing an explicit time-bounded execution record. The phone summarizes completed and skipped executions to support reflection and recipe revision.

This design intentionally prioritizes short, repeated, context-tethered actions. It is not intended for long-duration activities or actions that users routinely delay far beyond the trigger moment.

\subsection{Prototype Implementation} \label{sec:pebbl-implement}

We implemented \name as native iPhone and Apple Watch applications. Recipe authoring occurs on the phone through a chatbot interface that converts user input into a structured recipe representation. Trigger monitoring and execution confirmation occur on the watch. In the current prototype, trigger detection combines on-device sound classification, simple motion gating, and optional time/location constraints to reduce false activations and battery use. Once the user confirms execution, the watch records audio and inertial measurement unit (IMU) data until the user indicates completion, after which the files and associated metadata are transferred to the phone for storage.

The implementation prioritizes on-device inference for trigger detection and temporally sparse recording for uploaded execution data. Full implementation details, prompt design, UI flows, and detector logic are provided in Appendix~\ref{append:implementation}.

\section{Ethics, Consent, and Data Handling} \label{sec:ethics}

All study procedures were reviewed and approved by our university Institutional Review Board under No.~\texttt{03-2025-1005}. All participants provided informed consent before taking part and were informed about what data would be collected in each study, including audio, motion, location, interview recordings, and, in Study~2, neck-worn video. Participants could withdraw at any time without penalty, and could request deletion of their data.

Because \name{} involves privacy-sensitive sensing, we separated on-device sensing used for trigger detection from data retained for analysis. Trigger detection relied on on-watch audio and IMU processing, while non-triggered sensor streams were not uploaded to lab servers. For study analysis, we retained only confirmed execution records, including short audio clips, triaxial IMU, timestamps, recipe labels, and relevant time/location context. In Study~2, neck-worn video was used only as a reference for post-hoc annotation and was deleted within six months of study completion.

\begin{table*}[t]
\centering
\caption{Overview of the three studies. Per-study detail is in the corresponding \textit{Participants} subsections.}
\label{tab:study_overview}
\footnotesize
\setlength{\tabcolsep}{3.2pt}
\renewcommand{\arraystretch}{1.08}

\begin{tabularx}{\textwidth}{@{}
L{0.7cm}
L{1.05cm}
L{3.2cm}
L{1.75cm}
Y
L{3.05cm}
@{}}
\toprule
\textbf{Study} & \textbf{N} & \textbf{Participants} & \textbf{Setting} & \textbf{Purpose} & \textbf{Key outcomes} \\
\midrule
S1 & 6 
& 4M/2F, ages 23--27; wearable-HAR researchers with IMU data collection experience, from 4 research institutions 
& Expert workshop ($\sim$150 min) 
& Elicit evaluation criteria and expert assessment of \name relative to common data-collection workflows 
& Rubric dimensions, comparative ratings, qualitative tradeoffs \\

\addlinespace[2pt]
S2 & 21 
& 10M/11F, ages 21--31; 15 students, 4 early-career professionals, 2 others 
& Controlled in-lab study ($\sim$120 min) 
& Measure reliability of \name's execution records and compare perceived burden with alternative annotation workflows 
& Log recall/precision, wrong-trigger counts, temporal alignment, subjective ratings \\

\addlinespace[2pt]
S3 & 8 (3 overlap with S2) 
& 5M/3F, ages 20--27; own iPhone and Apple Watch users 
& Pilot deployment ($\sim$4\,h of \name use) 
& Assess whether the interaction and sensing pipeline can function outside the lab and how users perceive it in everyday use 
& Wrong-trigger rate, manual-fallback use, qualitative feedback, subjective ratings \\
\bottomrule
\end{tabularx}
\end{table*}

\section{Exploratory Study with Expert Workshop} \label{sec:study1}
We evaluate \name through three studies, summarized in \autoref{tab:study_overview}: an expert workshop (this section), a controlled in-lab study (Section~\ref{sec:study2}), and a pilot deployment in free-living settings (Section~\ref{sec:study3}). To understand \name's potential as a scalable data collection platform, we first consulted wearable-HAR researchers, who represent downstream data consumers familiar with the practical tradeoffs of collecting and annotating activity datasets. The workshop provided a researcher-facing assessment of \name's perceived benefits, limitations, and tradeoffs relative to existing collection workflows, complementing the participant-facing evidence in Studies 2 and 3. This study addresses \textbf{RQ1}.

\subsection{Experiment Design}
\begin{figure}
    \centering
    \includegraphics[width=\linewidth]{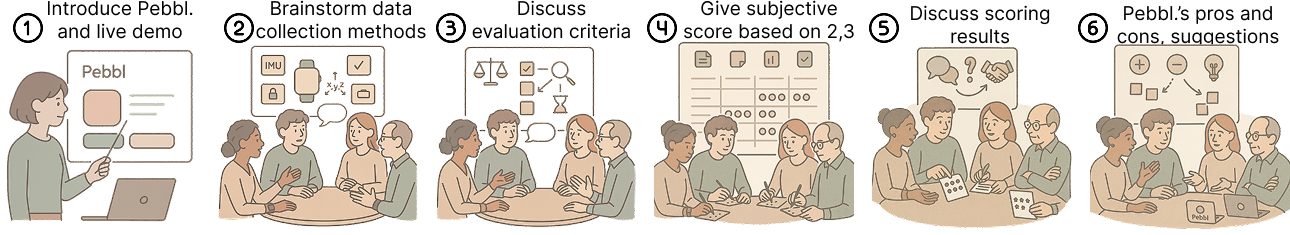}
    \caption{The procedure of the expert workshop (Study~1).}
    \label{fig:exp2_expert_workshop_pipeline}
\end{figure}

\subsubsection{Participants}

We recruited 6 wearable HAR researchers and individuals with vast experience in gathering wearable IMU datasets. The experts comprised 4 males and 2 females aged between 23 and 27 years old, and they came from 4 research institutions. Three of them were from academia, while the other three were affiliated with industrial research labs. They reported 2--6 years (mean $\approx$ 4) of experience in wearable sensing and/or activity dataset collection. The workshop lasted approximately 150 minutes, and the experts received compensation of 50 USD per hour.

\subsubsection{Procedure}

The workshop, illustrated in \autoref{fig:exp2_expert_workshop_pipeline}, had three goals: (1) identify the dimensions experts use to judge wearable activity data-collection workflows, (2) position \name relative to common alternatives, and (3) elicit likely strengths, limits, and deployment concerns. Participants first received an interactive demo of \name, including recipe generation and execution, and briefly tried the system on their own. We then presented three workflow families drawn from prior work: controlled protocolized collection~\cite{weiss2019wisdm, reiss2012introducing, chavarriaga2013opportunity}, free-living recording with post-hoc annotation~\cite{chan2024capture, grauman2022ego4d, xu2023towards}, and free-living recording with participant self-report~\cite{kim2022mymove, vaizman2018extrasensory} -- and asked experts to refine the evaluation criteria they considered most relevant when comparing these approaches with \name.

After agreeing on a rubric (i.e., the set of evaluation dimensions), experts independently rated each of the four approaches. We treat these ratings as expert judgments rather than objective performance measurements, because only \name was directly experienced during the workshop while the comparison approaches were evaluated based on prior experience and the literature above. A moderated discussion then surfaced disagreements and key tradeoffs, with one team member leading the conversation while another flagged dimensions with the widest rating spread for follow-up.

\subsubsection{Analysis Plan}

For each rubric dimension, we compare the scores experts gave to each data collection approach.
The qualitative data were analyzed by following \citet{braun2023doing}'s reflexive thematic analysis procedure; more details are provided in Appendix~\ref{append:exp_interview_braun}.

\subsection{Results and Findings} \label{sec:results_expert}

We report the workshop results in four parts: the data-collection approaches and rubric dimensions the experts converged on, the comparative scores they assigned to each approach, a summary of the overall assessment, and the qualitative feedback that complements those scores.

\subsubsection{Data Collection Methods \& Rubric Metrics}
The experts converged on the following list of wearable HAR data-collection approaches for our discussion:

\begin{itemize}
  \item \textbf{Predefined-class, Protocolized Collection in a Controlled Environment (M1):} Target activity categories are fixed and predefined; recruited participants wear sensing devices and follow a prescribed protocol to record these classes.
  \item \textbf{Open-world, Free-living with Post-hoc Annotation (M2):} Participants free-roam wearing both a recording device and a ``ground-truth'' device; afterwards, annotators label the data segments offline.
  \item \textbf{Open-world, Free-living with Participant Self-report (M3):} Participants free-roam wearing the recording device and label data either during the activity or shortly thereafter (within the day).
  \item \textbf{Ours (\name):} Our approach is similar to M3, but differs by pairing prospective self-report with in-situ confirmations, and by providing personal value through situated reminders for beneficial actions.
\end{itemize}

The dimensions by which the experts evaluated these approaches are defined in \autoref{tab:results-expert-criteria}.
Note that some of the items like \textit{Accuracy} and \textit{Bias} included sub-items for more depth.
The experts were instructed to score each method on a scale from 1–7.
For all dimensions other than those related to \textit{Cost} and \textit{Bias}, higher scores were more preferable.

\begin{table}[t]
    \centering
    \caption{
    The rubric dimensions used by experts to evaluate wearable HAR data-collection approaches. Note that the order in which these metrics are presented is weakly associated with their importance, as weighed by the experts.
    }
    \label{tab:results-expert-criteria}
    \resizebox{0.98\textwidth}{!}{%
    \begin{tabular}{llp{14cm}}
\toprule
\textbf{Primary Dimension} & \textbf{Sub-Dimension} & \textbf{Relevant Indicators and Measures} \\
\midrule
\multicolumn{2}{c}{\multirow{2}{*}{\textbf{Vocabulary Openness \& Coverage}}}  & Distinct action classes; proportion of fine-grained vs. coarse labels; coverage of verb–object–context combinations; distinct language annotation for similar activities; \\
\hdashline
\multicolumn{2}{c}{\textbf{\multirow{2}{*}{Scale \& Diversity}}} & Number of participants; recording length; environment/demographic diversity; per-activity instance counts; total annotated clips;  \\
    \hdashline
\multirow{5}{*}{\textbf{Accuracy}} & \textbf{Label Granularity} & Semantic specificity of labels (task-, object-, limb-level); presence of hierarchical labels; target task granularity; \\

& \textbf{Synchronization} & Estimated clock drift across devices; presence of alignment anchors (e.g., beeps, taps) \\
& \textbf{Accuracy of Label Timing} & Onset/offset jitter; missed/extra events rate; proportion of truncated/merged segments \\
& \textbf{Accuracy of Label Semantics} & Inter-annotator agreement on descriptions; contradiction rate across modalities \\
& \textbf{\multirow{2}{*}{Ecological Validity}} & Degree to which data reflects naturalistic behavior; fraction of free-living data; prompt-induced behavior rate; overlap with everyday routines; \\
\hdashline
\multirow{2}{*}{\textbf{Cost \& Feasibility}} & \textbf{Data Capture Difficulty} & User and researcher's burden; onboarding/training time; hardware requirements; failure/retry rate; \\
& \textbf{Data Annotation Difficulty} & Minutes of annotation per hour of data; need for expert annotators; instruction complexity; throughput; \\
\hdashline
\multirow{3}{*}{\textbf{Bias}} & \textbf{Activity Bias} & Class imbalance (long-tail); coverage of rare/complex actions; diurnal/weekday bias \\
& \textbf{Participant Bias} & Age, gender, handedness, ability, culture; recruitment channels; exclusions \\
& \textbf{Context Bias} & Scene diversity; seasonality/weather coverage; \\
\hdashline
\multicolumn{2}{c}{\textbf{Privacy Protection}}  & Strength of privacy-by-design: data minimization, on-device processing, encryption, user transparency; \\
\hdashline
\multicolumn{2}{c}{\multirow{2}{*}{\textbf{Multi-modality Support}}} & Number of modalities (e.g., accel, gyro, photoplethysmography, audio); sampling rates; missing-data rate; multi-device fusion support \\
\bottomrule
    \end{tabular}
    }
\end{table}

\begin{table*}[t]
\scriptsize
\centering
\setlength{\tabcolsep}{2.2pt}
\renewcommand{\arraystretch}{1.12}
\caption{The average scores of 1-7 Likert Scale given by the experts when evaluating different wearable HAR data-collection approaches. The arrows next to the column names indicate whether higher ($\uparrow$) or lower ($\downarrow$) scores are better. The best scores in each column are highlighted in bold.}
\label{tab:results_expert_subjective}

\resizebox{\textwidth}{!}{%
\begin{tabular}{l cc ccccc cc ccc cc}
\toprule
\multirow{2}{*}{\textbf{Method}} &
\multicolumn{2}{c}{\textbf{Coverage$\uparrow$}} &
\multicolumn{5}{c}{\textbf{Accuracy$\uparrow$}} &
\multicolumn{2}{c}{\textbf{Cost \& Feasibility$\downarrow$}} &
\multicolumn{3}{c}{\textbf{Bias (diversity)$\downarrow$}} &
\textbf{Privacy$\uparrow$} &
\textbf{Sensors$\uparrow$} \\
\cmidrule(lr){2-3}\cmidrule(lr){4-8}\cmidrule(lr){9-10}\cmidrule(lr){11-13}\cmidrule(lr){14-14}\cmidrule(lr){15-15}
& \textbf{Vocab.} & \textbf{Scale}
& \textbf{Granu.} & \textbf{Synchro.} & \textbf{Timing} & \textbf{Semantics} & \textbf{Validity}
& \textbf{Capt. Diff} & \textbf{Ann. Diff}
& \textbf{Act. Bias} & \textbf{User Bias} & \textbf{Ctx Bias}
& \textbf{Privacy} & \textbf{Sensors} \\
\midrule
M1   & 1.50 & 2.33 & 4.67 & \textbf{6.33} & \textbf{6.50} & \textbf{6.50} & 2.00 & 4.83 & 3.33 & 5.67 & 5.33 & 6.17 & 4.67 & \textbf{5.83} \\
M2   & 4.67 & 4.67 & \textbf{5.50} & 5.17 & 5.67 & 5.83 & 5.50 & 4.33 & 6.00 & \textbf{3.00} & 3.83 & 3.50 & 1.67 & 5.00 \\
M3   & 5.50 & 5.83 & 3.50 & 4.67 & 3.50 & 4.50 & 5.67 & 3.83 & 4.33 & 3.33 & 3.00 & 2.33 & 6.17 & 3.17 \\
Ours & \textbf{6.67} & \textbf{6.50} & 4.83 & 5.00 & 5.00 & 5.50 & \textbf{6.50} & \textbf{2.33} & \textbf{2.33} & 3.17 & \textbf{2.67} & \textbf{2.17} & \textbf{6.33} & 3.17 \\
\bottomrule
\end{tabular}%
}
\end{table*}

\subsubsection{Summary of Expert Assessment}
\autoref{tab:results_expert_subjective} summarizes the experts' ratings. We interpret these scores as an elicitation of expert expectations about likely tradeoffs, not as definitive comparative evidence: only \name was directly experienced during the workshop, whereas the three comparison workflows were rated from prior experience and the literature. With that caveat, experts generally saw \name as promising on \textit{Vocabulary Openness}, \textit{Scale}, \textit{Ecological Validity}, \textit{Capture} and \textit{Annotation Difficulty}, and \textit{Privacy Protection}, while viewing controlled protocolized collection (M1) as stronger on \textit{Synchronization}, \textit{Timing}, and \textit{Semantic Accuracy}, and post-hoc annotation (M2) as stronger on \textit{Label Granularity}. \name\ and M3 trailed on \textit{Multi-modality Support}, reflecting a deliberate tradeoff favoring deployability and privacy over maximal sensing. Experts also emphasized that \name's advantages are contingent on recipe fit: the approach is most plausible for short, repeatable, context-linked activities, and less plausible for actions that are delayed, socially awkward, or poorly captured by the current trigger set.

\subsubsection{Qualitative Feedback}

\paragraph{\textbf{Pros.}} Expert~5 appreciated that \name could minimize unintended data collection by only recording short windows of data at moments when users previously agreed to be recorded. Expert~1 commented that \name would impose \textit{``almost no user burden''}; meanwhile, Expert~6 went as far as saying that people may forget that they are collecting data for research rather than just scaffolding a healthy habit, which may results in collecting more natural behavior. Expert~3 stated, \textit{``The tiny-habit premise tends to attract self-motivated users. They might be more responsible; more detailed,''}. They later hypothesized that the annotations generated by such individuals would be more reliable.

\paragraph{\textbf{Cons.}} Expert~3 worried that by focusing on beneficial activities, less positive or hedonic activities (e.g., gaming, smoking) would be difficult to capture using \name. Expert~1 had similar concerns by noting, \textit{``It would probably only cover adolescents to middle-aged; older adults usually wouldn't use it.''} These two experts also expressed concerns about the precise timing of the recordings. Expert~3 worried that users may start the recording after seeing the trigger notification on their watch, only to complete the activity when they find a more convenient opportunity, while Expert~1 worried that some users may forget to stop recording upon activity completion. Expert~1 also expressed concerns that extraneous or poorly timed triggers may annoy users.

\paragraph{\textbf{Design Trade-offs.}} The experts highlighted a tension between \name's detailed onboarding and its intention of promoting an open vocabulary of activities. Expert~3 cautioned that prescriptive guidance risks narrowing what users create: \textit{``The design threshold is high. Some users may struggle to imagine how to design recipes ~\dots If you give especially good examples, everyone may just use those.''} On the other hand, Expert~1 noted that if \name's guidance was too loose or generic, users would have no clear direction on how they should use the system. Expert~4 further warned that \textit{``Without feasibility cues, end users may propose undetectable actions, such as opening some desktop app, that wearable sensors can't capture''}.

\paragraph{\textbf{Suggestions for Future Refinements.}} The experts suggested several system refinements for future work. Although we did not implement these recommendations in this work, we provide them here to inspire later iterations:
\begin{enumerate}
\item Broaden and automate the trigger space to include modalities beyond audio and automatically discover salient cues for future recipes (Experts~2, 4, and 6).
\item Add automated verification of action completion to prevent a \textit{``false sense of accomplishment when users confirm but do not actually execute an action''} (Expert~6).
\item Enrich the agent's dialogue to help users design recipes for breaking undesirable habits (Expert~1).
\item Allow the system to work with only a smartphone to enable more engagement (Expert~1). Expert~5 remarked that implementing this idea might introduce new challenges, saying \textit{``We don't know where the phone is and positions change across activities''}.
\end{enumerate}

\section{Main Study: In-Lab Evaluation} \label{sec:study2}
To support whether the experts' perspectives on \name would materialize in practice, we invited several individuals to evaluate it as both users and data annotators.
This study addresses two research questions: ``How efficient and reliable is \name as an approach to scalable data collection?'' (\textbf{RQ2}), and ``Can \name encourage people to take beneficial actions during everyday living?'' (\textbf{RQ3}).

\subsection{Experiment Design}
\begin{figure}
    \centering
    \includegraphics[width=\linewidth]{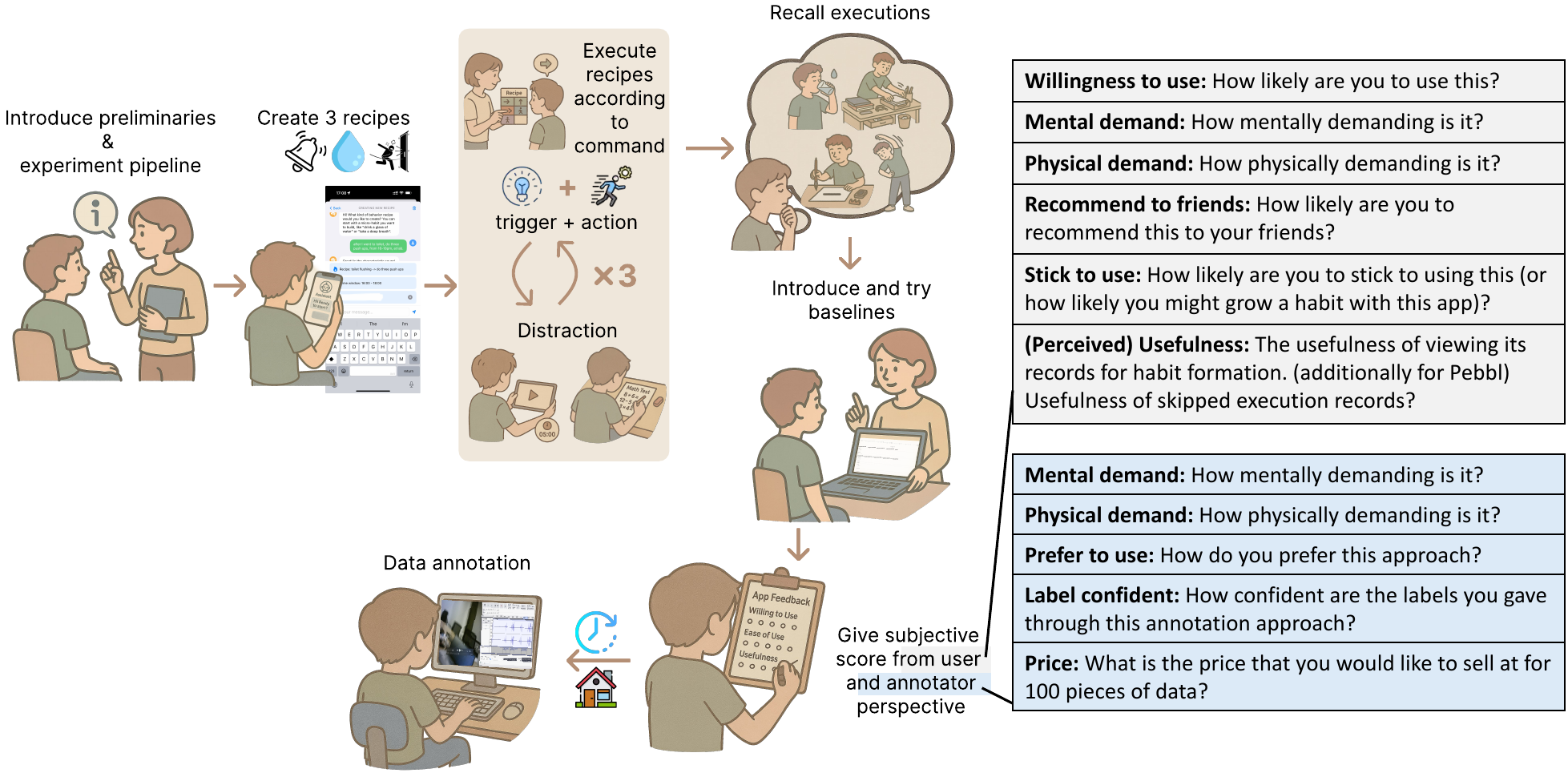}
    \caption{The procedure of the in-lab user study (Study~2).}
    \label{fig:exp2_user_study_pipeline}
\end{figure}

\subsubsection{Participants}

We included 21 participants, comprising 10 males and 11 females aged between 21 and 31 years old. Participants were recruited from a community-based pool of adult smartphone users, yielding 15 university students, 4 early-career professionals, and 2 individuals from other occupational backgrounds; \textit{$15$} reported prior use of a smartwatch, and \textit{$5$} reported prior use of a habit-tracking or self-logging app. The experiment lasted approximately 120 minutes, and they received compensation of 30 USD per hour. 

\subsubsection{Comparison Conditions}

Study~2 included two classes of comparison. First, to evaluate \name as a data-labeling workflow, we compared it against three alternatives that represent common annotation strategies discussed in Study 1: post-hoc video annotation (\textbf{B1}), near-time self-report (\textbf{B2}), and an in-situ phone-based self-report interface inspired by prior work (\textbf{B3}). Second, to evaluate the user-facing reminder/tracking aspect of \name, we compared it against a commercial habit-tracking app (\textbf{B4}). \autoref{tab:study2_baselines} summarizes these conditions and the outcomes for which each was used. We do not claim that these baselines exhaust the design space; rather, they provide representative contrasts for annotation burden, label confidence, temporal quality, and user-facing tracking preference.

\begin{table*}[t]
\centering
\small
\caption{Study~2 comparison conditions. B1--B3 are annotation-workflow comparisons; B4 is used only for the user-facing reminder/tracking comparison. Detailed implementations and citations are given in the text below.}
\label{tab:study2_baselines}
\begin{tabular}{p{0.6cm}p{3.0cm}p{5.0cm}p{4.6cm}}
\toprule
\textbf{ID} & \textbf{Represents} & \textbf{Participant task} & \textbf{Compared on} \\
\midrule
B1 & Post-hoc video annotation & Review neck-camera video and assign labels and temporal boundaries using an annotation tool & Annotation burden, confidence, and temporal quality \\
\addlinespace
B2 & Near-time self-report & Record activities shortly after performing them & Annotation burden and confidence; comparison to retrospective memory \\
\addlinespace
B3 & In-situ self-report with phone UI & Label activities using a mobile interaction similar to prior self-report systems & Annotation burden and confidence \\
\addlinespace
B4 & Commercial habit tracker & Manually log completed actions using a publicly available tracking app & User-facing reminder/tracking preference only \\
\bottomrule
\end{tabular}
\end{table*}

For the annotation-workflow comparison, the three baselines were implemented as follows:
\begin{enumerate}
\item \textbf{B1}~\cite{grauman2022ego4d, chan2024capture, wang2024dreamcatcher}: Annotators examined time-aligned ground-truth video recorded by a neck-worn camera and gave text labels to selected intervals using a professional labeling tool\footnote{https://www.audacityteam.org}.
\item \textbf{B2}~\cite{kim2022mymove}: Annotators labeled events shortly after recording them.
\item \textbf{B3}~\cite{vaizman2018extrasensory}: We implemented a demo similar to the ExtraSensory app\footnote{\url{https://github.com/cal-ucsd/ExtraSensoryAndroid}} to reflect a fine-grained open-vocabulary interaction paradigm.
\end{enumerate}
Examples of these interfaces are provided in Appendix~\ref{append:impl_baseline}.

For the user-facing tracking comparison, we used a publicly available habit-tracking app\footnote{https://apps.apple.com/us/app/habit-tracker/id1438388363} (\textbf{B4}).
This app offers a counter that users can increment either immediately after completing a target action or later in the day after recalling its execution.

\subsubsection{Procedure}

The study was held in a meeting room positioned less than 10~meters from a washroom, a kitchen, an elevator, and the building's waste handling area, meaning that there was moderate background noise throughout the entire experiment.
\autoref{fig:exp2_user_study_pipeline} summarizes the procedure, which can be divided into four phases:

We designed Phase~2 (the main \name condition) as a controlled measurement rather than a naturalistic use trace. Triggering each recipe on a pre-generated schedule, with the experimenter randomly instructing execution or skipping, yields known ground truth against which \name's records can be validated. We therefore interpret the reliability numbers reported below as evidence about the interaction's correctness under controlled conditions; the free-living pilot (Study~3) provides the naturalistic complement.

\begin{enumerate}
    \item \textbf{Introduction:} After introducing participants to the concept of tiny-habit recipes, we asked them to brainstorm three recipes with distinct audio triggers (examples can be found in Appendix~\ref{append:exp_recipe_examples}). We then provided them with materials that explained \name and the experiment procedure; these materials can be found in Appendix~\ref{append:exp_intro}.
    \item \textbf{Main Experiment:} Participants were asked to trigger each of their recipes 30 times. Using a monitor, a research team member instructed participants to either skip or complete the corresponding action at random. After the 7$^\text{th}$ and 16$^\text{th}$ trials, participants were distracted by a 5-minute entertainment video or arithmetic problems to disrupt any habituation between trials. Throughout this session, participants were required to wear a camera around their neck, serving as our ground truth for annotation.
    We collected two sets of egocentric videos with different camera configurations: one with electronic image stabilization (EIS) enabled and one without. We did this because most popular smart-glass camera models ship without built-in EIS~\cite{engel2023project, baumann2023neon}, so both settings are realistic deployment scenarios for the baselines.
    \item \textbf{Post-experiment:} Participants were asked to recall how many times they actually completed the actions associated with each of the triggers, providing an indicator of whether participants could rely on their recent memory for annotation. We then allowed participants to experiment with the data-annotation and behavior-tracking baselines for a few trials, after which they provided feedback from the perspectives of both users and annotators. Because \name{} was evaluated as a full end-to-end workflow with extensive trials, whereas the baselines were introduced through short hands-on trials to support comparison, we interpret these ratings as participants' informed perceptions of the relative burden, confidence, and preference associated with each workflow.
    \item \textbf{At-home Annotation:} After participants left the lab, we sent them the video captured by the camera they wore during the study so that they could provide baseline annotations for executed actions. To ease some of the annotation burden, we provided them with approximate execution times noted by the research team member so that they could quickly locate relevant segments. Videos were 60-80 minutes long, and participants spent 30-60 minutes annotating them. Every submission was reviewed by the research team, anchored against the pre-generated execution list from Phase~2 to catch missing entries, duplicates, and obvious mislabels. This was a lightweight consistency check to make sure that participants fully understand the annotation procedure.
\end{enumerate}

\subsubsection{Analysis Plan}

Study~2 focuses on \textit{controlled reliability}, not long-term adoption. We therefore analyze four objective outcomes --- (1) whether \name's execution records match the known trial schedule, (2) how well participants retrospectively remember executions, (3) how often the trigger detector produces false prompts, and (4) how \name's time boundaries compare with post-hoc video annotation --- together with subjective ratings used to assess perceived burden and confidence for representative comparison workflows. Long-term adherence and sustained naturalistic use are out of scope for this study; the pilot deployment (Study~3) addresses those questions at a preliminary scale.

The pre-generated execution order list served as ground-truth labels for when actions were completed.
Meanwhile, the camera served as ground-truth information for the actions' timestamps.
Due to camera issues mid-study, we were only able to analyze camera feed data from 18 participants: 9 recorded with EIS and 9 without. 
To align \name's clock with the camera's, we scanned the camera's video with a 0.1-second step length to find the latency that maximized the intersection-of-union (IoU) between participants' annotations and \name's records. 
This information allowed us to conduct objective evaluations to assess three aspects of \name: (1) the accuracy of its resulting records, (2) the rate of incorrect triggers, and (3) the precision of its resulting time boundaries. 

\begin{itemize}
    \item \textbf{Label Accuracy:} We report the recall, precision, and F1 score of \name's exported execution records relative to the pre-generated execution order list from the main experiment. True positives are defined as instances when a \name record is as defined in the order list, false positives are defined as instances when a \name record is not as defined in the order list, and false negatives are defined as instances when an execution in the order list is not recorded by \name.
    \item \textbf{Participant Memory:} We define the recall of participants' memory as one minus the mean absolute error (MAE) between recalled and ground-truth execution counts, normalized by the total number of executions: $1-\frac{\text{MAE}(\#executions_{gt}-\#executions_{recall})}{\#executions_{total}}$. $\#executions_{gt}$ refers to the ground truth provided by the generated execution list. We further analyzed participants' recall for successfully executed recipes and skipped recipes, as well as execution during each session. We included this metric to assess how accurately participants can retrospectively recall their own executions, given that both our in-the-wild data collection baseline and the baseline tracking app rely on after-the-fact labeling.
    \item \textbf{Wrong Trigger Rate:} We counted the \texttt{Wrong Trigger} clicks for each category and each participant, treating them as user-reported false positives when \name wrongly recognized an irrelevant ambient event as the recipe trigger. 
    \item \textbf{Temporal Alignment:} We calculate the  IoU, temporal recall/precision, and length ratio of annotated actions. 
\end{itemize}

The subjective evaluation metrics listed in \autoref{fig:exp2_user_study_pipeline} were scored along a 7-point Likert scale, except for price. Because \name provides extra incentive by offering reminder + tracking features, we explicitly require users to take that into account when evaluating the price.

Our post-study questionnaire combined adapted workload items (mental demand and physical demand from NASA-TLX) with custom items targeting willingness to use, recommendation likelihood, stickiness, perceived usefulness, annotation preference, and label confidence. These custom items were authored by the research team to probe the adoption and annotation constructs most relevant to \name's two comparison classes (reminder/tracking vs.\ annotation workflow). We used subsets of the items in each comparison --- the user-facing comparison used the reminder/tracking subset; the annotation-workflow comparison used the annotation subset. The full list of items and the exact wording administered to participants are reproduced in Appendix~\ref{append:study2_questionnaire}.

\subsection{Results and Findings}

We report objective measures of \name's reliability (label accuracy, wrong-trigger occurrence, participant memory, and temporal alignment), followed by participants' subjective ratings under the two comparison classes: the reminder/tracking-side comparison and the annotation-workflow comparison.

\subsubsection{System Performance Evaluation.}
\begin{table}[t]
\caption{\name's labeling accuracy.}
\label{tab:results_inlab_accuracy}
\resizebox{0.98\textwidth}{!}{%
\begin{tabular}{ccccc|cccc}
\toprule
\textbf{Recall} & \textbf{Precision} & \textbf{F1} & \textbf{Recall} (success) & \textbf{Recall} (skip) & \textbf{P. recall} & \textbf{P. recall} (success) & \textbf{P. recall} (skip) & \textbf{P. recall} (sessions) \\
\midrule
97.30\% & 97.15\% & 97.22\% & 98.06\% & 96.64\% & 70.32\% & 76.73\% & 61.71\% & 51.75\% \\
\bottomrule
\end{tabular}
}
\end{table}

\paragraph{\textbf{Controlled reliability of \name's execution logs.}} \autoref{tab:results_inlab_accuracy} shows that \name's execution records closely matched the pre-generated trial schedule under controlled conditions (recall $=97.30\%$, precision $=97.15\%$, F1 $=97.22\%$). This result reflects the strength of \name's interaction design: the user is prompted at the trigger moment and must explicitly confirm execution or skipping. We therefore interpret this finding narrowly as evidence that \name can produce reliable execution logs in a structured setting. It should not be read as proof that all resulting data windows are equally valid in unconstrained real-world use, where users may delay actions, miss prompts, or stop recordings late.

\paragraph{\textbf{Retrospective recall was substantially noisier.}} Participants' retrospective counts were markedly less reliable than \name's in-situ records. Average memory accuracy was $70.32\%$ overall, with better recall for successful executions than skipped ones ($76.73\%$ vs.\ $61.71\%$). Session-level recall dropped to $51.75\%$, indicating difficulty placing events accurately in time even over a short study period. Wilcoxon signed-rank tests at the experiment level were significant at $p<10^{-5}$ across all comparison pairs. These results support the core motivation for \name: when logging is deferred, participants reconstruct rather than report, and that reconstruction is error-prone.

\begin{figure}
  \centering
  \includegraphics[width=0.4\linewidth]{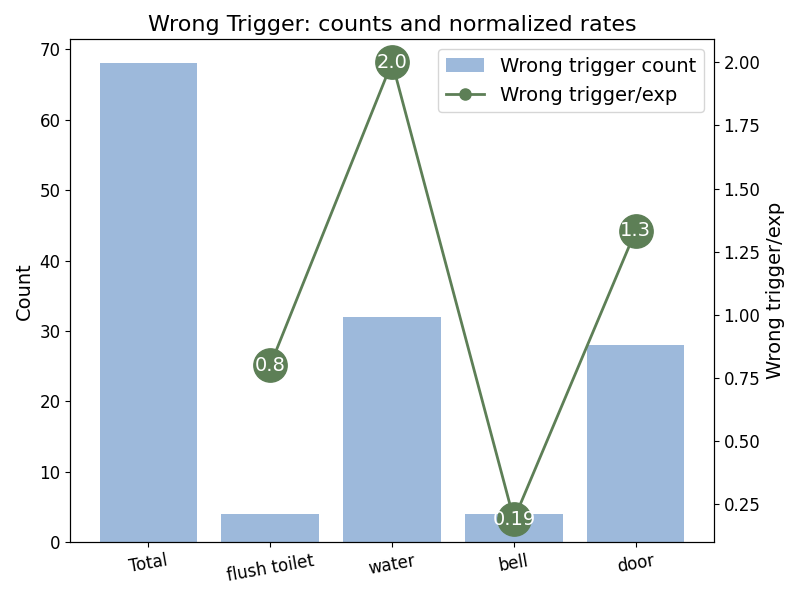}
  \caption{Wrong-trigger counts and per-experiment rates by trigger class. The per-participant breakdown is in Appendix~\ref{append:wrong_trigger_per_participant}.}
  \label{fig:eval_wrong_trigger_count}
\end{figure}

\paragraph{\textbf{False triggers remained a practical limitation.}} \label{sec:exp2_wrong_trigger_analysis}
Across 630 true trigger opportunities, participants reported 68 wrong-trigger events ($10.79\%$ relative to true triggers; see \autoref{fig:eval_wrong_trigger_count}). After normalizing by the number of experiments exposed to each trigger class, errors concentrated in \textit{water} ($\approx 2.0$ wrong/exp) and \textit{door} ($\approx 1.3$ wrong/exp), with \textit{flush toilet} intermediate ($\approx 0.8$) and \textit{bell} most robust ($\approx 0.19$). This pattern is consistent with acoustic distinctiveness: bells and toilet flushes have more stereotyped temporal--spectral signatures, whereas water and doors are acoustically diverse and frequently co-occur with domestic confounders. Per-participant variation (Appendix~\ref{append:wrong_trigger_per_participant}) was concentrated in a small tail driven by the same two noisy classes and by proximity to a washroom, kitchen, and waste area less than $10$\,m from the lab. Deployment quality therefore depends strongly on acoustic distinctiveness and on the ambient context in which each recipe is active.

\begin{figure}
    \centering
    \includegraphics[width=\linewidth]{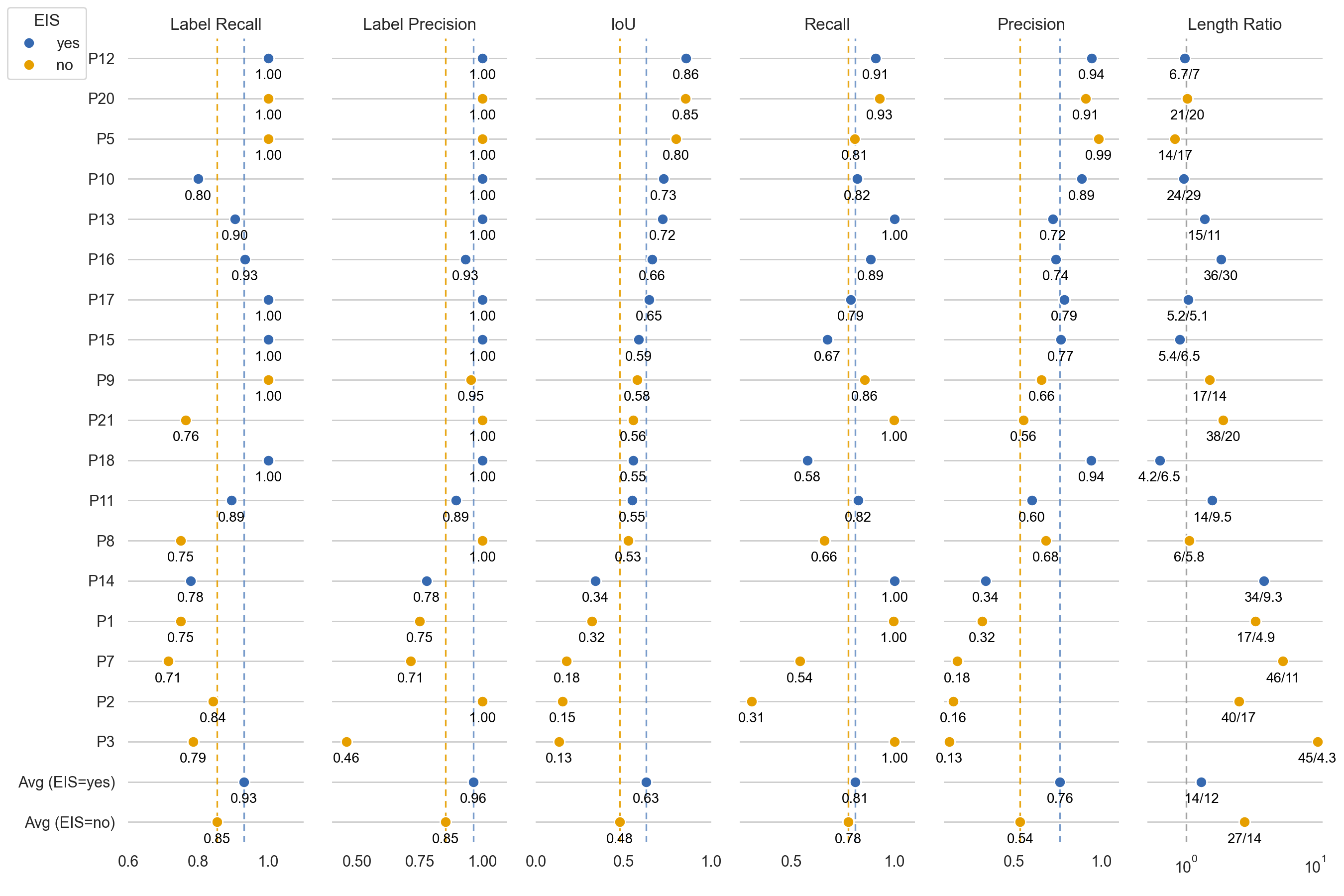}
    \caption{Annotation Quality Analysis. We show results of both description-level and segment-level evaluation metrics per participant. Orders were rearranged by annotation IoU, which is the primary evaluation metric. ``Length Ratio'' represents the average interval length of the participant's annotations over \name's recordings. We draw the vertical line of Length Ratio $=1$ in the last column instead of the average value. Electronic Image Stabilization (EIS) represents two camera types used in our experiment. }
    \label{fig:eval_annotation_acc}
\end{figure}

\paragraph{\textbf{Time boundaries were tighter than post-hoc annotation, but still instruction-sensitive.}} Compared with post-hoc video annotation, \name generally produced tighter temporal windows (\autoref{fig:eval_annotation_acc}): most participants show high recall but low precision when labeling video, with length ratios $>1$, indicating that annotators tend to over-segment events, whereas \name's Start/End taps anchor the recording around the execution. The EIS condition yields somewhat tighter averages than the non-EIS condition, but per-participant variability dominates device effects, as is common in fine-grained temporal annotation. This comparison depends on participants following the intended interaction pattern: in Study~2 they were explicitly instructed to tap \texttt{Start} just before acting and \texttt{End} immediately after completion. Real-world use is likely to be messier, so we treat this result as evidence about the usefulness of the interaction design rather than a guarantee of boundary fidelity in deployment. Section~\ref{sec:discuss_deploy_quality} discusses mitigations.

\subsubsection{User-Facing Perceptions of \name as a Reminder/Tracking Tool}

\begin{figure}
    \centering
    \includegraphics[width=\linewidth]{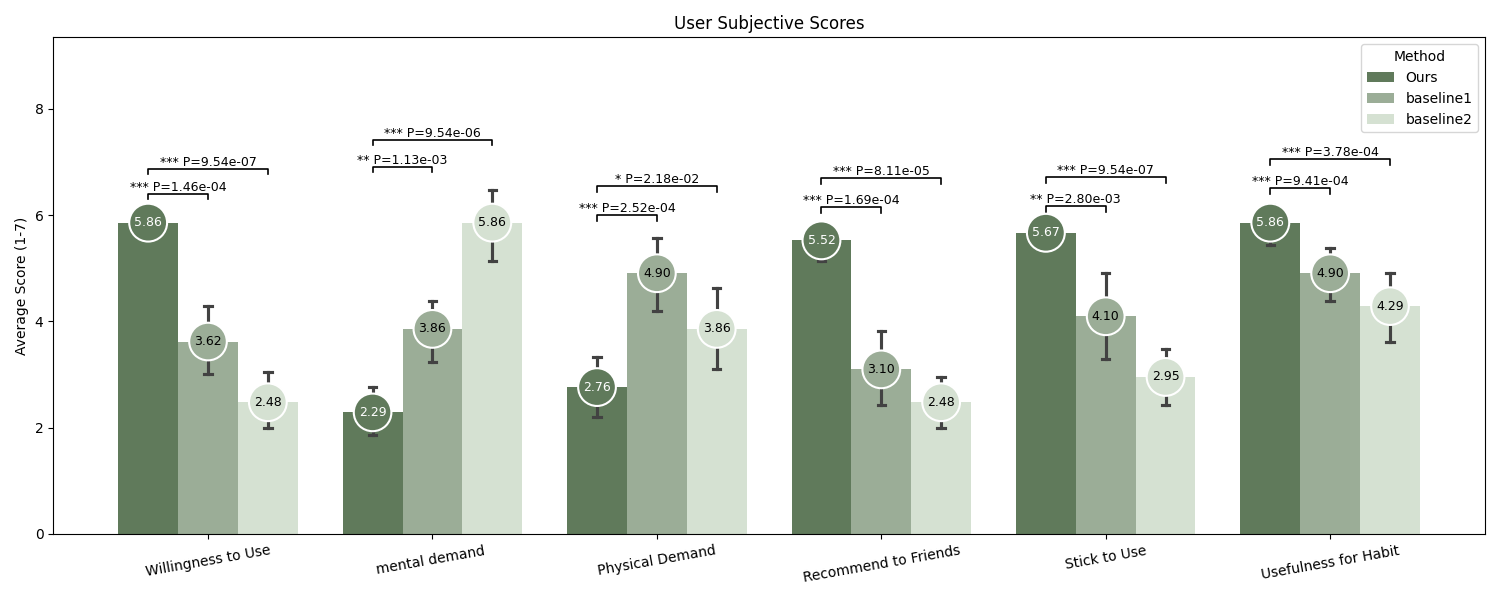}
    \caption{Subjective ratings under the user-facing reminder/tracking comparison (\name vs.\ B4 and the two self-report--style conditions). Per-dimension means, signed-rank $p$-values, and effect sizes are reported in the figure; ** denotes $p<0.01$ and *** denotes $p<0.001$.}
    \label{fig:eval_user_subjective}
\end{figure}

\autoref{fig:eval_user_subjective} summarizes participants' ratings for the user-facing tracking comparison. Under the representative exposure used in Study~2 --- brief trials with the comparison conditions relative to the full \name session --- participants reported higher willingness to use \name, lower mental and physical demand, higher likelihood of recommending it, higher likelihood of continued use, and greater perceived usefulness of \name's records for reflection. Exact means and $p$-values (Wilcoxon signed-rank) are reported in the figure. Interview comments suggest that the main driver was not visual design but timing: \name asks for confirmation at the activity opportunity, whereas the baseline relies on later manual logging. Participants also valued the visibility of \textit{skipped} executions ($5.48\pm0.85$ on a 1--7 scale), which several described as useful for reflecting on why they did not act and for revising the recipe accordingly rather than just ``trying harder''.

We interpret these ratings cautiously. Study~2 does not establish that \name improves long-term adherence or habit formation; it shows that, during short-term controlled exposure, participants perceived the reminder-plus-logging interaction as more supportive than a purely manual tracker. Study~3 provides a preliminary free-living view of the same perceptions.

\subsubsection{Perceived Burden and Confidence as an Annotation Workflow}

\begin{figure}
    \centering
    \includegraphics[width=\linewidth]{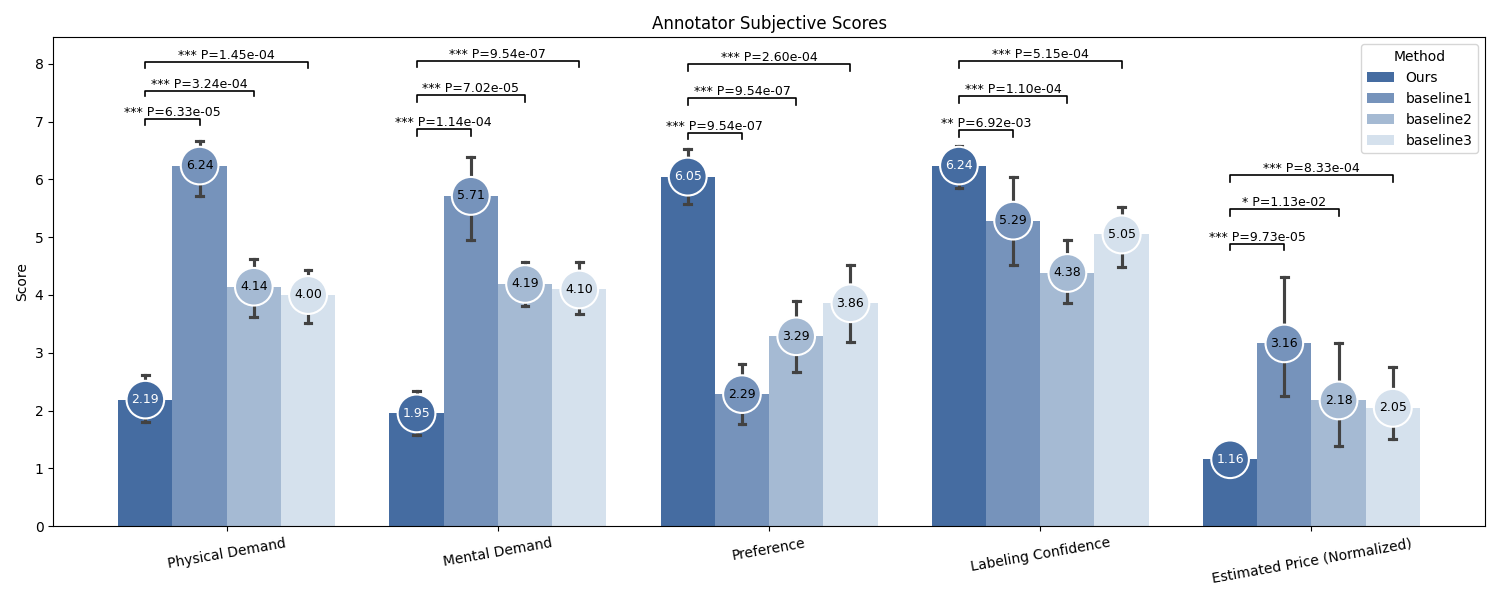}
    \caption{Subjective ratings under the annotation-workflow comparison (\name vs.\ B1--B3). Per-dimension means, signed-rank $p$-values, and effect sizes are reported in the figure; ** denotes $p<0.01$ and *** denotes $p<0.001$. The \textit{price} dimension originally included in this figure has been relocated to Appendix~\ref{append:study2_price}.}
    \label{fig:eval_annotator_subjective}
\end{figure}

\autoref{fig:eval_annotator_subjective} summarizes participants' ratings of the annotation-oriented comparison conditions (B1--B3). Participants consistently rated \name as lower in mental and physical demand than B1--B3 and expressed higher confidence in the labels produced through \name; exact means and $p$-values (Wilcoxon signed-rank) are reported in the figure. This pattern is consistent with the system design: \name collapses collection and labeling into a single in-situ confirmation at the trigger moment, so annotators do less retrospective reconstruction and fewer label-choice/policy-doubt decisions than under B1--B3. The confidence gap is particularly notable against B1 (post-hoc video annotation): in Study~1, experts judged video-based annotation's \emph{label quality} as strong, yet participants here reported lower \emph{production-side} confidence. The two measurements are not in conflict --- they capture different constructs (theoretical quality under aggregate review vs.\ experienced confidence while producing a label).

We do not interpret these ratings as evidence that \name universally dominates all annotation workflows. Rather, for short, triggerable activities under the representative exposure used in Study~2, participants experienced \name as a less effortful way to produce bounded, time-anchored labels than retrospective or phone-centric alternatives. Whether that advantage transfers to longer-duration activities, less distinctive triggers, or sustained use remains open.

\section{Pilot Deployment Study in Open-Environment} \label{sec:study3}
Since the scenarios in the previous study were largely contrived, we conducted a pilot deployment to further support our answers to \textbf{RQ2} and \textbf{RQ3}, which respectively speak to \name's efficacy as a data collection tool and a way of promoting beneficial actions. We frame this deployment explicitly as a pilot: it assesses whether \name's interaction and sensing pipeline can function outside the lab, whether its specific failure modes (wrong triggers, context mismatch) reappear in free living, and how a small group perceives \name as a reminder/tracking tool during short-term use. It is not intended to establish sustained adoption, long-term retention, or behavior change.

\subsection{Experiment Design}

We describe the pilot's participants, the additional implementation work needed to support free-living deployment, and the procedure that participants followed.

\subsubsection{Participants}
We recruited 8 participants (3 overlap with Study~2), comprising 5 males and 3 females, aged from 20 to 27 ($24\pm2.73$). Participants were required to have an iPhone and an Apple Watch themselves. Participants were \textit{college students}, and $6$ reported prior use of a habit-tracking or self-logging app. The experiment lasted approximately 10 hours, with 120 minutes of actual execution time, and they received compensation of 70 USD in total.

\subsubsection{Additional Implementation}
\begin{figure}[t]
    \centering
    \includegraphics[width=0.8\linewidth]{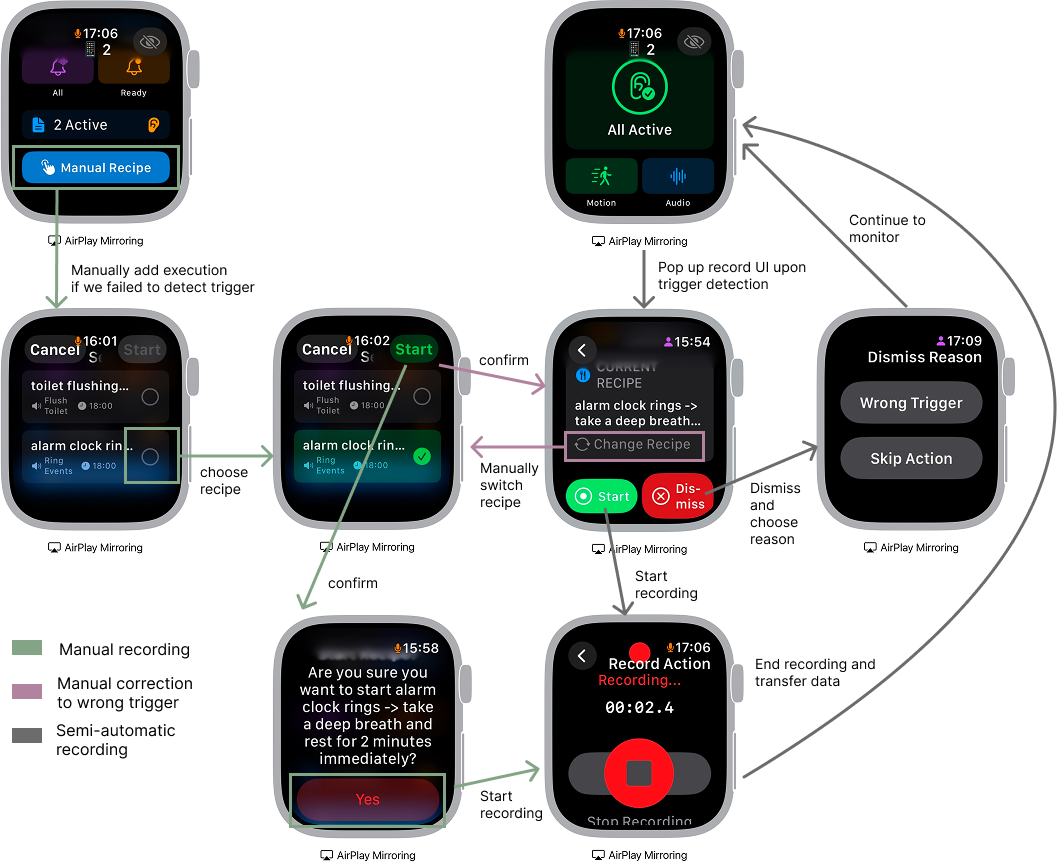}
    \caption{In-the-wild on-watch interaction flow. Compared to Section~\ref{sec:pebbl-implement}, we add two manual operations.}
    \label{fig:itw_implement}
\end{figure}

We add manual recording approaches compared to Section~\ref{sec:pebbl-implement}, as shown in~\autoref{fig:itw_implement}. Specifically, users can manually start execution recording by clicking \texttt{Manual Recipe}. Meanwhile, users can click on \texttt{Change Recipe} to choose the correct recipe to execute if \name detects a trigger incorrectly, or the user missed the haptic feedback of the last wrong trigger.

\subsubsection{Procedure}
In-the-wild study comprised four sessions:

\begin{enumerate}
    \item \textbf{Introductory Session:} Participants check and update system version beforehand. After they come to the lab, we invite them to our testing group and install \name to their devices through TestFlight\footnote {\url{https://developer.apple.com/testflight/}}. Similar to Study~2, we first introduce participants to the concept of tiny habits and then ask them to design one example recipe. Then we introduce the usage of \name and the study pipeline. Afterwards, we confirm the location and time range that participants want to try \name within, then they brainstorm two recipes to execute for the study. 
    \item \textbf{\name Session:} Participants were required to try out \name for approximately 4 hours, with 2 active recipes in a free-living context. Each recipe should be triggered approximately 4 times, which is a much sparser distribution compared to Study~2. This lower density was intentional: the pilot is designed to reflect intermittent natural use rather than the repeated scripted execution used in Study~2, and therefore trades statistical volume for a more deployment-like trigger schedule. Within the location and time range, participants stick to their original schedule and perform recipe triggers intermittently. After the tryout, they send the execution records and recorded data to us.
    \item \textbf{Baseline Session:} Participants were required to try out a baseline app\footnote{\url{https://apps.apple.com/us/app/habit-tracker/id1438388363}} for a similar time length, same recipes, location, and free-living context, as well as triggering recipes approximately 4 times. 
    \item \textbf{Post-experiment Session:} We conduct an online scoring and interview session afterwards. The evaluation metrics are identical to user perspective metrics in Study~2 (see ~\autoref{fig:exp2_user_study_pipeline}). During scoring, we explicitly require participants to focus on functionality and weaken the impact of the UI's artistic design, which is not our emphasis.
\end{enumerate}

\subsection{Results and Analysis}
The results below should be read as \emph{pilot} evidence. They come from a short ($\sim$4 hours of \name use), small ($N=8$), and technically narrow sample, and are not designed to support claims about long-term retention, sustained engagement, or behavior change at scale. What the pilot can show is (i) whether \name's interaction and sensing pipeline functions in free living, (ii) whether the failure modes we observed in the lab (wrong triggers, context mismatch) reappear under intermittent natural use, and (iii) how a small group's Study~2-style subjective perceptions survive a short baseline comparison outside the lab. Limitations of the sample and duration are revisited in the Discussion.

\subsubsection{Objective Analysis of In-the-Wild Reliability for \name}

\begin{figure}
  \centering

  \subcaptionbox{Environment.\label{fig:itw_exp_context}}
    {\includegraphics[width=0.3\linewidth]{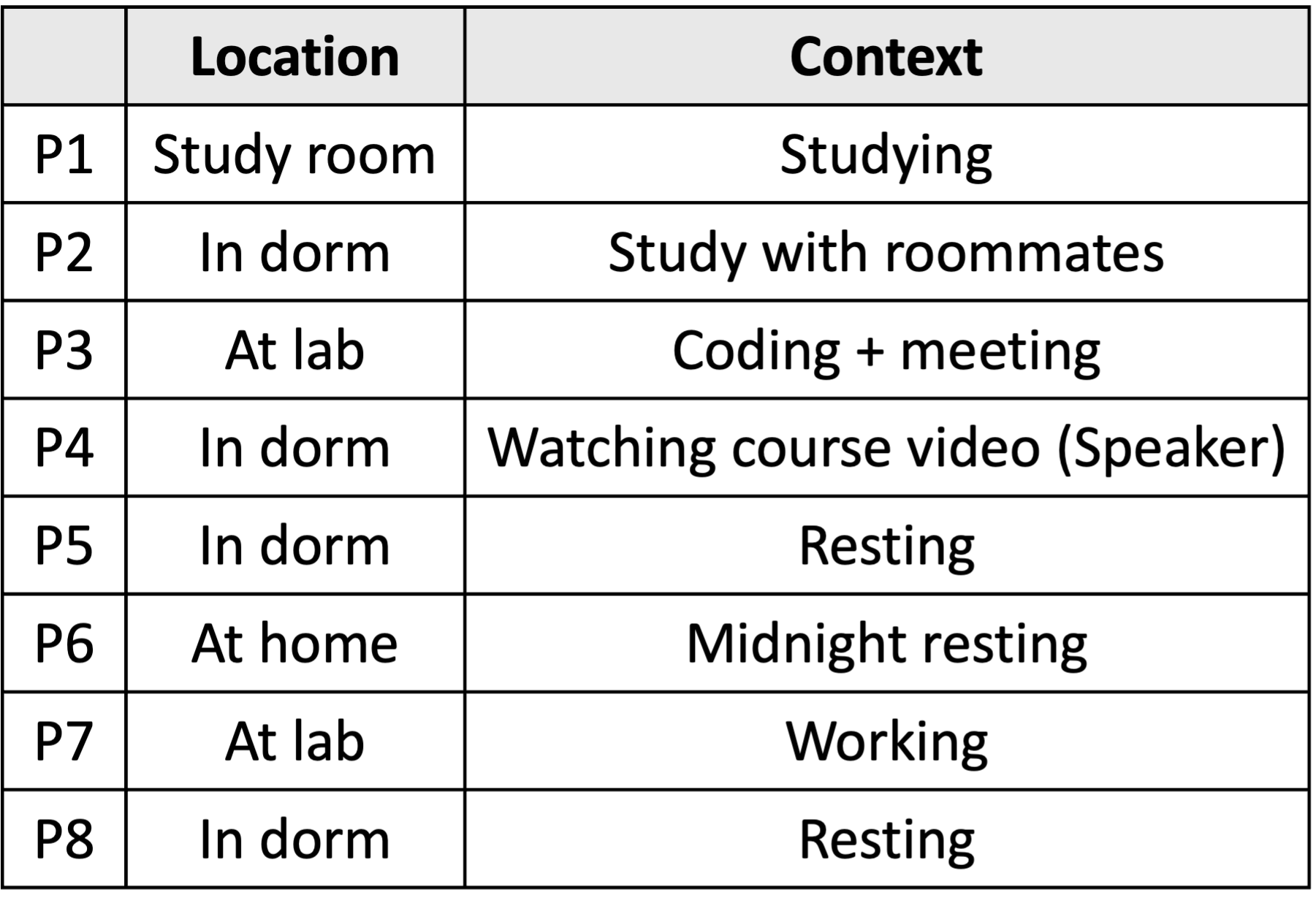}}
  \subcaptionbox{Wrong trigger and manual recipe rate.\label{fig:itw_wrongtrigger_participant}}
    {\includegraphics[width=0.6\linewidth]{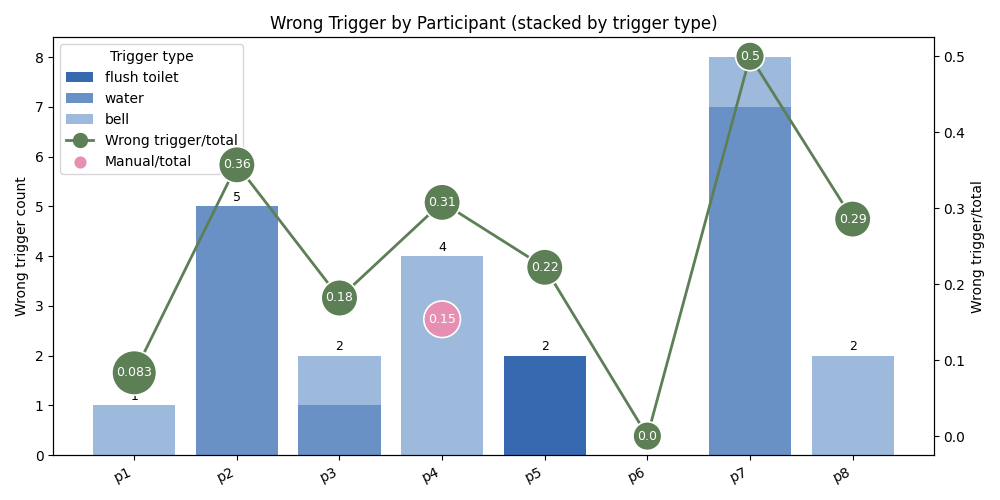}}
  
  \caption{Experiment information and performance per participant.}
  \label{fig:itw_info_eval}
\end{figure}

We count the wrong trigger rate and the manual recipe rate for each participant. Different from Study~2, we calculate both rates by dividing the total record count, as participants no longer execute triggers identically in this study. As shown in~\autoref{fig:itw_info_eval}, \name produced a moderate and context-dependent wrong trigger rate. P2 suffers from a frequent wrong triggering of the water recipe due to the fan's noise, while P4 reported that the video course's sound effect played through the speaker can sometimes arouse the ringtone recipe. P7 experienced a frequent water recipe triggering whenever he was near the washroom. However, it is not very bothersome, as this kind of fine-grained location-aware error also serves as a strong reminder, according to P7 in the interview. Conversely, only P4 has manually recorded recipes for 2 times. This demonstrates that \name is relatively sensitive, which is noticed by most participants.


Taken together, Study 3 shows that wrong triggers are highly context-sensitive rather than uniformly distributed across use cases. Although wrong triggers occurred less frequently per hour than in the noisier lab environment, true executions were also much sparser in free-living use; therefore, each false prompt could be more noticeable and affect perceived burden. We therefore treat trigger robustness and context fit as central constraints for real-world deployment.

\begin{figure}
    \centering

    \subcaptionbox{Statistical tests.\label{fig:itw_subjective_avg}}
    {\includegraphics[width=0.49\linewidth]{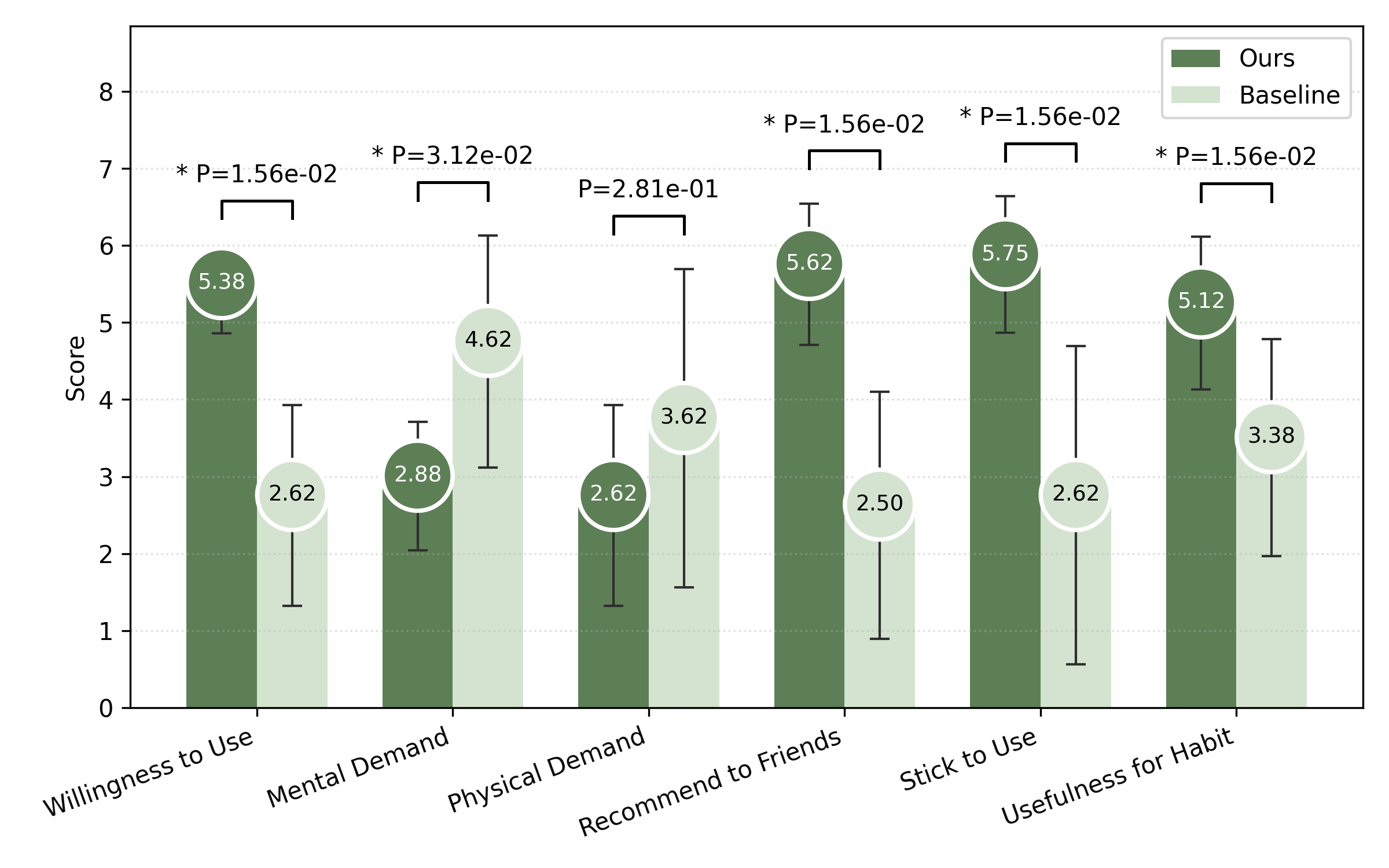}}
  \subcaptionbox{Descriptive statistic. Connection lines are paired scores from the same participant. \label{fig:itw_subjective_descriptive}}
    {\includegraphics[width=0.49\linewidth]{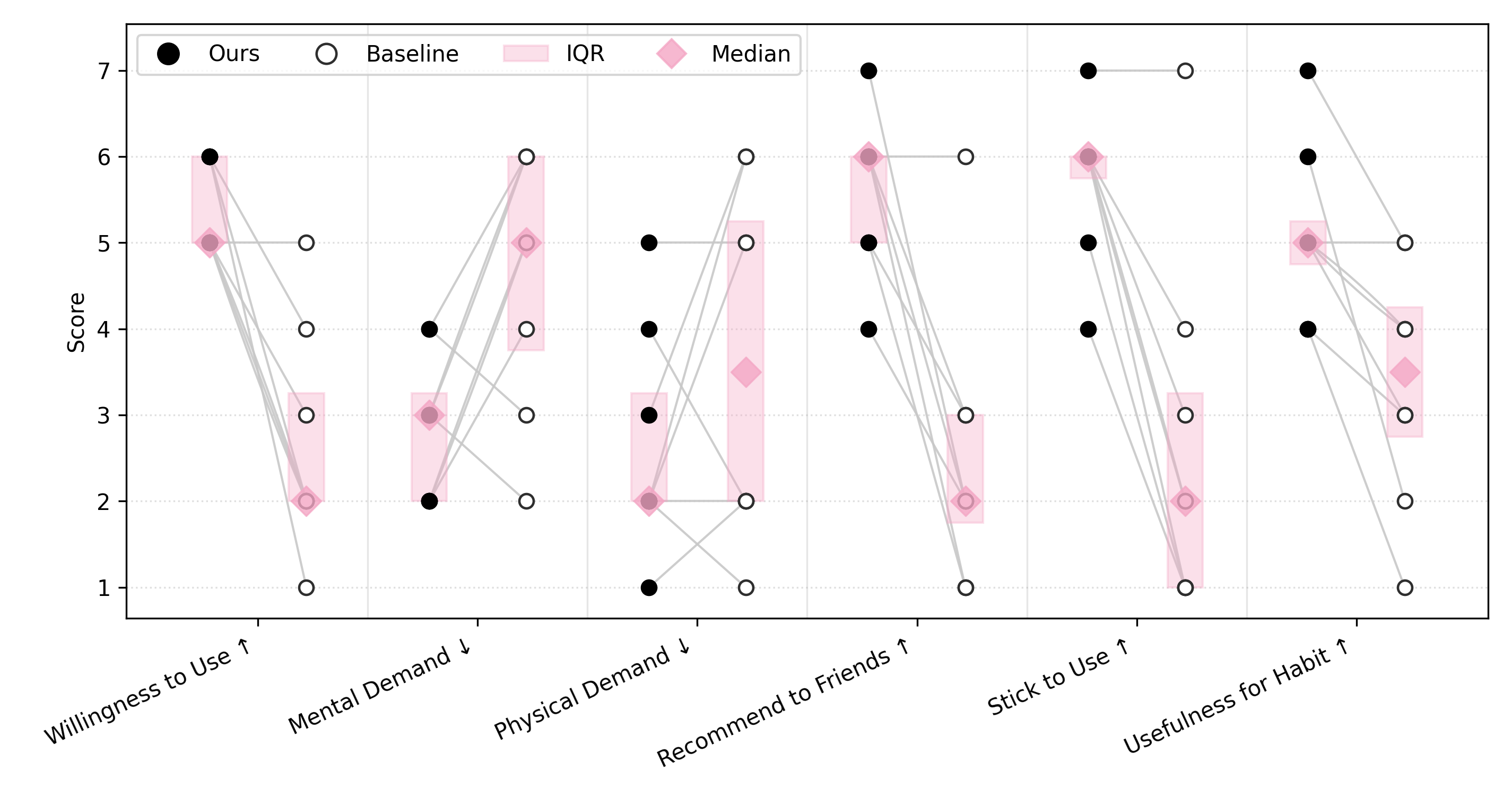}}
    \caption{In-the-wild subjective scores. (a) Per-dimension means: participants favored \name across all six dimensions, with significance indicated by Wilcoxon signed-rank tests ($p<0.05$ marked with *). (b) Per-participant scores shown as discrete markers (filled = \name, hollow = baseline), with the pink band indicating the inter-quartile range (IQR) and the pink diamond indicating the median.}
    \label{fig:itw_subjective}
\end{figure}

\subsubsection{In-the-Wild User Preference for \name.}
\autoref{fig:itw_subjective} shows participants' subjective scores for \name and the baseline, visualized as per-participant markers with median and IQR overlays. Participants generally gave higher scores for \name with respect to ``willingness to use'', ``stick to use'', ``recommend to friends'', and ``(perceived) usefulness''. These trends are consistent with Study~2. For mental demand, \name tended to be rated as less demanding than the baseline, although P6 stated, \textit{``When I was just starting the training, it felt a bit, um, complicated''}, indicating a steep learning curve during onboarding. Physical demand scores were more mixed. P2 and P4 regarded \name as equally burdensome to the baseline because of the relatively higher wrong trigger rate, while P6 and P8 regarded \name as more burdensome because \name can effectively nudge them to execute the beneficial actions: \textit{``Honestly, I think that burden is good. Without pressure, I won't start''} (P8). Taken together, these descriptive results corroborate Study~2 and suggest that \name delivers higher perceived value and adoption potential, at the cost of some additional initial effort.

\subsubsection{User-Centered Insights on \name's Usability and Future Directions.} \label{sec:study3_result_interview}
In this section, we summarize the qualitative feedback we received during our interviews. Participants' full responses can be found in Appendix~\ref{append:exp_interview_quote}. 

\paragraph{\textbf{Overall positive reception.}} Participants valued \name's ``smart'' automation (P8) and tight watch–phone integration (P4/5), which made capturing moments feel effortless and timely. The haptic feedback created a rewarding sense of interaction (P3). Several participants (P6/7/8) noted that \name's ``gentle pressure'' to act was actually productive, kick-starting beneficial actions that have the potential to form long-term habits. 

\paragraph{\textbf{Need for recipe generation guidance.}} For some participants (P2/3/7), \name had a steep learning curve because they needed more guidance to pick recipes that would be suitable for their daily living. They were unsure of which actions to target and which triggers to pick to create the strongest connection. Users asked for stronger support, such as a conversational agent that can iteratively convert vague intentions into concrete recipes.

\paragraph{\textbf{Suitability of audio-based triggers.}} 
Participants felt that audio-based triggers were a strong fit for short and repeatable behaviors such as drinking water, brief exercise sets, and breaking up sedentary time (P4/7). Longer actions that required more time were not viewed as suitable for \name due to the need for more specific contexts that could not be conveyed by time, location, and audio alone (P1/2). 
Therefore, some participants wanted richer and more personalized triggers beyond the current audio set.
Beyond supporting triggers according to fixed timers, other contextual variables included sedentary time (P2), screen time (P5/7), social media usage (P5/7), and other health-related routines like medication usage (P4). P7 also suggested that smarter feedback summaries that helped explain when triggers were activated would help them make potential adjustments.

Taken together, Study~3 shows that \name's pipeline functions in free-living settings and that its perceived-value signals from Study~2 are directionally reproduced under a brief baseline comparison, while also re-surfacing the onboarding friction and trigger-context limitations that experts flagged in Study~1. We interpret these findings as a pilot assessment of deployment feasibility and boundary conditions, not as evidence of sustained real-world effectiveness; we return to sample size, duration, and the longitudinal question in the Discussion.

\section{Discussion}

Across the three studies, we examine \name from both sides of the proposed data-collection approach: researchers who may use the resulting wearable activity data, and users who produce data through everyday interaction. For \textbf{RQ1}, Study 1 provides a researcher-facing assessment of \name's perceived value and tradeoffs relative to common wearable activity data-collection workflows. Experts saw \name as promising for vocabulary openness, scalability, ecological validity, low capture/annotation burden, and privacy-preserving data minimization. At the same time, they highlighted important limitations around trigger coverage, temporal fidelity, recipe fit, activity bias, and participant diversity.

\textbf{RQ2} and \textbf{RQ3} examine the user/data-provider side from two different angles: RQ2 focuses on whether \name can produce useful labeled data, while RQ3 focuses on whether the same interaction provides user-facing reminder/tracking value. We therefore studied both questions in Studies 2 and 3, because each \name interaction simultaneously produces a labeled sensor window and supports a tracking/reminder experience. Study 2 provides controlled evidence that \name can produce reliable execution logs, tighter temporal boundaries than post-hoc annotation, lower perceived annotation burden, and higher label confidence. Study 3 complements this evidence in free-living use, where participants' feedback was broadly consistent with Study 2 but false triggers and context-dependent recipe fit became more visible. Thus, our findings support \name as a feasibility-stage approach for low-burden, user-contributed wearable activity labeling, but not yet as evidence of long-term habit formation or sustained behavior change.

We now discuss the implications, scope, and deployment challenges in this section.

\subsection{Trigger-Action Tracking as a Wearable Labeling Paradigm}

The main contribution of \name{} is a trigger-action tracking paradigm for low-burden wearable data labeling. Rather than asking users to retrospectively reconstruct what they did, \name{} lets users define an intended action in advance, waits for a relevant trigger, and then asks for a lightweight confirmation at the moment the action is expected to occur. Across the expert workshop, controlled lab study, and pilot deployment, our findings suggest that this interaction can reduce perceived annotation burden while still producing timestamped execution records with user-authored, open-vocabulary labels.

The key design insight is that a trigger-action recipe can serve as a form of prospective self-report. In conventional self-report workflows, the user must observe, remember, and later describe an activity after it happens. In \name{}, the user first specifies a meaningful action -- for example, doing a short exercise, drinking water, or checking an object -- as part of a personally useful tracking routine. This action description becomes a prospective label and when the trigger occurs, the watch prompt turns this prospective label into an in-situ confirmation.

This reframing helps explain why \name{} was perceived as lower burden than retrospective alternatives. The system does not merely make annotation faster; it changes when and why annotation happens. First, it reduces the memory burden of labeling by surfacing the recording interface at a moment of opportunity, rather than asking users to search backward through time. Second, it reduces semantic ambiguity by asking users to define the target action before execution, instead of requiring later interpretation of opaque IMU or audio traces. Third, it gives the labeling interaction immediate personal value: the same confirmation that contributes data also helps users track, reflect on, or carry out actions they already care about. The broader implication is that wearable data collection can be embedded into lightweight, personally valuable routines, rather than framed only as collection and annotation labor.

\subsection{Application Scope and Boundary Conditions}

Study~3 suggests that the current prototype is better suited to short, repeatable actions that can be performed soon after a recognizable cue. This boundary comes partly from the current implementation: although users can describe actions in open-vocabulary language, the deployed prototype is configured around four audio trigger classes. Thus, the scope of suitable habits depends on the available trigger space. Expanding the audio vocabulary or adding system-, motion-, physiological-, and context-based triggers could broaden the range of habits \name{} can support, but such expansion is a deployment direction rather than a claim evaluated in this paper.

This application scope also shapes the data \name{} can collect. Because \name{} embeds labeling into habit-tracking routines, it is more likely to capture actions that users perceived beneficial. This produces a biased sample of all everyday activities. Undesired behaviors, privacy-sensitive actions, and behaviors from users who are less likely to adopt smartwatch-based habit tools may be underrepresented. We therefore frame \name{} as a mechanism for low-burden in-situ labeling within a bounded class of beneficial routines, rather than as a general-purpose solution for all types of activity. This bounded scope is also population-dependent: recipes that are appropriate for our young, technically comfortable participants may not transfer directly to older adults, children, or users with mobility/accessibility needs. For such groups, \name may need different task domains (e.g., medication routines, light mobility, hygiene, or caregiver-approved chores), stronger recipe-authoring assistance, and targeted validation of both usability and data quality.

\subsection{Future Work: Toward At-Scale Real-World Deployment}
\label{sec:discuss_deploy_quality}

Our studies suggest that trigger-action tracking is a promising interaction paradigm, but moving from a feasibility-stage prototype to public deployment requires additional design and engineering work. We discuss three directions: broadening the trigger space, improving data quality safeguards, and supporting everyday onboarding and retention.

\textbf{Broadening, personalizing, and composing richer trigger contexts.} The current prototype uses audio cues as a tractable first instantiation, but many behaviorally meaningful situations have no clear acoustic signature. Future versions should support multimodal triggers, including physiological signals such as Heart-Rate Variability (HRV), device-usage signals such as screen-time thresholds or app events, and motion/context cues such as prolonged sedentary periods or posture. Trigger detection should also become more personalized: few-shot calibration, adjustable thresholds, and per-user trigger examples could help adapt detection to individual routines, environments, and devices. Not all useful triggers are single events. Real-world routines often depend on combinations of time, location, recency, motion, and device state. Future systems could support compound trigger templates such as ``after my last bathroom visit at night'' or ``after sitting for two hours and then standing up,'' allowing users to express more meaningful contexts while preserving the lightweight trigger-action structure.

\textbf{Maintaining data quality at scale.} Public deployment will introduce data-quality challenges that were only partially visible in our studies. Users may start or stop recording imprecisely, perform a related but different action, or use semantically vague labels. Rather than adding heavy verification gates, future systems should use lightweight safeguards: onboarding trials for temporal boundary fidelity, optional practice examples as weak references for execution filtering, simple trim/discard tools, and agent-assisted recipe creation to encourage more specific labels. Post-hoc language-model clustering may further help normalize open-vocabulary labels for downstream analysis.

\textbf{Supporting onboarding and long-term use.} Participants valued the intelligent prompting loop, but Study~3 also showed that recipe authoring can be unfamiliar. Future deployments should therefore balance scaffolding with openness: beginner templates, inspiration galleries, conversational recipe refinement, and adaptive feedback could help users start without collapsing the diversity of user-authored actions. Longer-term feedback mechanisms, such as progress summaries, streaks, badges, or supportive reflections on skipped executions, are also important for studying whether \name{} can sustain engagement over weeks or months.

\subsection{Limitations}
\label{sec:limitations}

Our findings should be interpreted within several limitations. First, the participant samples were relatively young, technically comfortable, and situated within the Apple ecosystem. Second, the studies were short-term: Study~2 provides controlled evidence and Study~3 offers only a short pilot deployment in participants' own contexts. We therefore cannot claim long-term retention, sustained motivation, or habit formation. A longitudinal deployment is needed to understand recipe stability, annoyance from false triggers, and whether the tracking interaction remains valuable over time. Third, the current studies do not produce a dataset large enough to validate downstream open-vocabulary modeling utility. Our results support the feasibility of collecting bounded, user-labeled sensor windows, but they do not demonstrate that such data improves activity-recognition models. Future work should evaluate downstream modeling once deployment-scale data are available. Finally, \name{} is currently implemented on iOS/watchOS, and we did not conduct a systematic multi-device battery benchmark. Although the implementation uses on-device inference, motion-gated audio processing, geofence-gated activation, and explicit sensor cleanup to reduce energy cost, deployment-scale claims about battery life require controlled evaluation across devices and watchOS versions.

\subsection{Privacy Considerations}
\label{sec:discuss-privacy}

\begin{figure}[!t]
    \centering
    \includegraphics[width=1\linewidth]{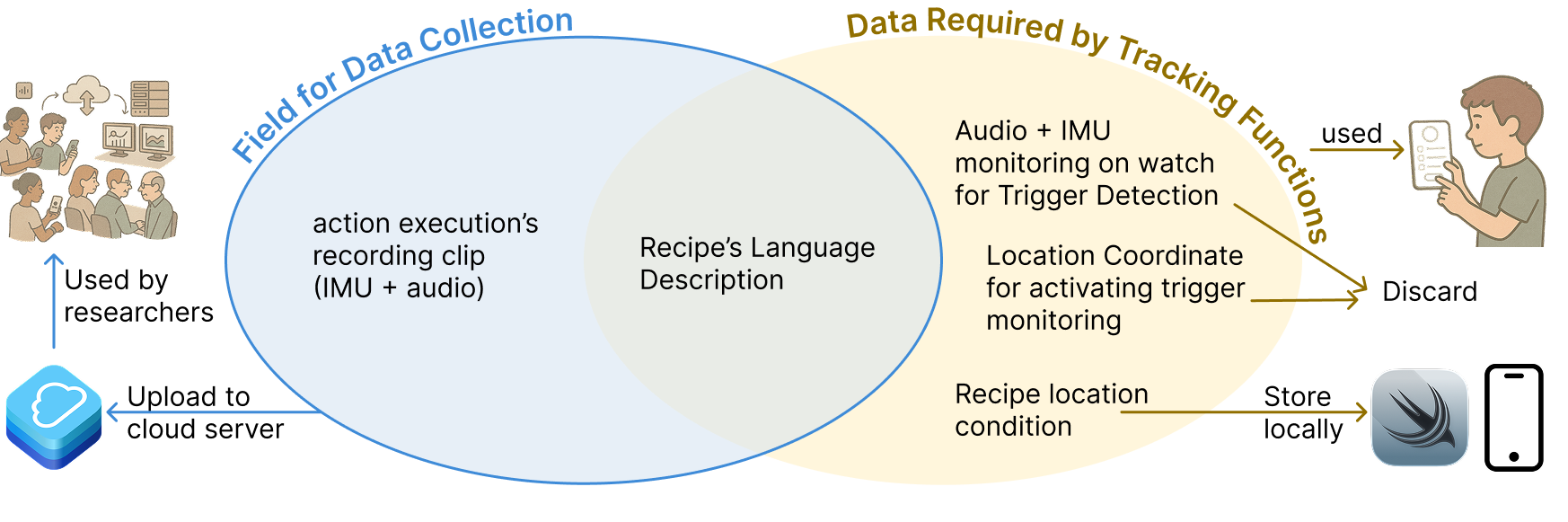}
    \caption{Data usage in \name. Continuous audio, IMU, and location streams are processed on-device for trigger detection and then discarded; only confirmed execution windows (short audio/IMU clips plus the user-authored action description) leave the device for analysis.}
    \label{fig:discuss_privacy}
\end{figure}

\name depends on privacy-sensitive sensing, especially audio and location (\autoref{fig:discuss_privacy}). Our design reduces exposure by performing trigger detection on-device and retaining only temporally sparse execution windows rather than continuous recordings; the ethics and data-handling procedures for the three studies are covered separately in Section~\ref{sec:ethics}. Even with on-device processing, participants may perceive always-on listening as intrusive, and some activities or locations may simply be inappropriate for this sensing model. The persistent microphone indicator on the watch also communicates ``being recorded'' regardless of whether raw data actually leaves the device, which can itself undermine feelings of safety. Expanding beyond audio-based triggers is therefore not only a technical opportunity but also a privacy requirement for broader adoption; future trigger modalities must continue to respect the minimization principles adopted here.

\section{Conclusion}

We presented \name, a smartphone--smartwatch system that uses user-authored trigger-action routines to support in-situ labeling of wearable activity data. In the current prototype, triggers are limited to four audio-based classes (door closing, toilet flushing, running water, ringtone/alarm), while actions are described in open-vocabulary natural language. Across an expert workshop ($N=6$), a controlled in-lab study ($N=21$), and a pilot deployment ($N=8$), we found that \name can reduce perceived labeling burden and produce reliable execution logs (recall $=97.30\%$, precision $=97.15\%$) \emph{under controlled conditions}, while also exposing practical constraints around trigger coverage, onboarding, and deployment context that we treat as part of the paper's scope rather than as minor implementation details.

Taken together, these results position \name as a promising feasibility-stage approach to user-centered wearable activity labeling. More broadly, they suggest that wearable data collection may become more sustainable when labeling is embedded in interactions that users find personally useful rather than purely burdensome; whether that sustainability materializes at scale, over longer time horizons, and with more diverse populations remains an open question for future deployment studies.

\section{LLM Use Clarification}

Human characters in Figures~\ref{fig:teaser}, ~\ref{fig:methods_user_journey}, ~\ref{fig:exp2_expert_workshop_pipeline} and ~\ref{fig:exp2_user_study_pipeline} are generated by GPT-5. For paper writing, we use GPT-5 for grammar correction and sentence refinement. In our implementation, we use GPT-4o to build a chat agent for creating recipes. 

\begin{acks}
    This work is supported by the National Key R\&D Program of China under Grant No. 2024YFB4505500.
\end{acks}

\bibliographystyle{ACM-Reference-Format}
\bibliography{sample-base}

\appendix
\newpage
\section{Implementation Details}
\label{append:implementation}

This appendix reports implementation details that complement the main paper, including the recipe-creation agent prompt, the UI and interaction flows, the baseline conditions used in Study~2, the audio-trigger-choice rationale, and the end-to-end system architecture and detection pipeline.

\subsection{Prompt}
The prompt for recipe creation chat agent:

\begin{lstlisting}
You are an intelligent behavior-aware assistant that helps users record real-life behaviors as habits and micro-habit triggers to build their own "behavior recipes". Your responsibilities are as follows:

1. Understand the characteristic sound corresponding to the trigger condition described by the user, which should be one of the following four: door closing, ringing, toilet flushing, or other water sounds.
2. Add only one piece of information per round to guide the user to gradually complete the recipe, do not ask multiple questions at once.
3. If the micro-habit recipe provided by the user meets the requirements, gently guess the approximate time period the user wishes to perform the micro-habit recipe (e.g., "daytime" or "nighttime") and ask the user for confirmation.
4. Further confirm with the user the specific time interval for executing the micro-habit recipe, which should be in whole hours.
5. Friendly guide the user to confirm the location where the micro-habit recipe occurs (e.g., "home" or "other places").
6. Among the two limiting conditions of time and place, the user must provide at least one explicitly.
7. Do not use the word "remind". Instead, use phrases like "recorded", "your habit has been saved", or "added to your habit list" to emphasize recording and support, rather than simple reminders.
8. Technical analysis is only reflected in the backend fields imu_desc, audio_desc, and condition, and is not mentioned to the user in the reply.

Your questions are in order:
1. Confirm the characteristic sound corresponding to the trigger condition with the user.
2. Confirm the approximate time period for executing the micro-habit recipe with the user.
3. Confirm the specific time interval for executing the micro-habit recipe with the user.
4. Confirm the location where the micro-habit recipe occurs with the user.

Please reply in the following JSON format, all fields must be provided:
- reply: Natural language guidance or confirmation for the user (within two sentences).
- recipe:[trigger, action] Behavior recipe, where trigger is the trigger condition and action is the behavior. Both are strings. If the user has not provided them yet, return an empty string "". Connecting the trigger and action with a comma should form a coherent sentence. Use the user's original words as much as possible.
- imu_desc: Description of the backend analysis of IMU detectability (return an empty string "" if there is no relevant information).
- audio_desc: Type of characteristic sound: Close Door, Ring Events, Flush Toilet, or one of the Water Events. Please analyze which of the above four types the sound provided by the user is. Return an empty string "" if there is no relevant information. If it can correspond to one of the four, record the corresponding English name.
- time_condition: Time condition, format is [start_hour, end_hour] array, e.g., [8, 22] for 8 AM to 10 PM (return an empty array [] if not set).
- location_condition: Location description (return an empty string "" if there is no relevant information).
- finished: Boolean value, true when all conditions are clear and the behavior is detectable.
\end{lstlisting}

\subsection{Detailed Interaction and UI}

\begin{figure}
    \centering
    \includegraphics[width=1.0\linewidth]{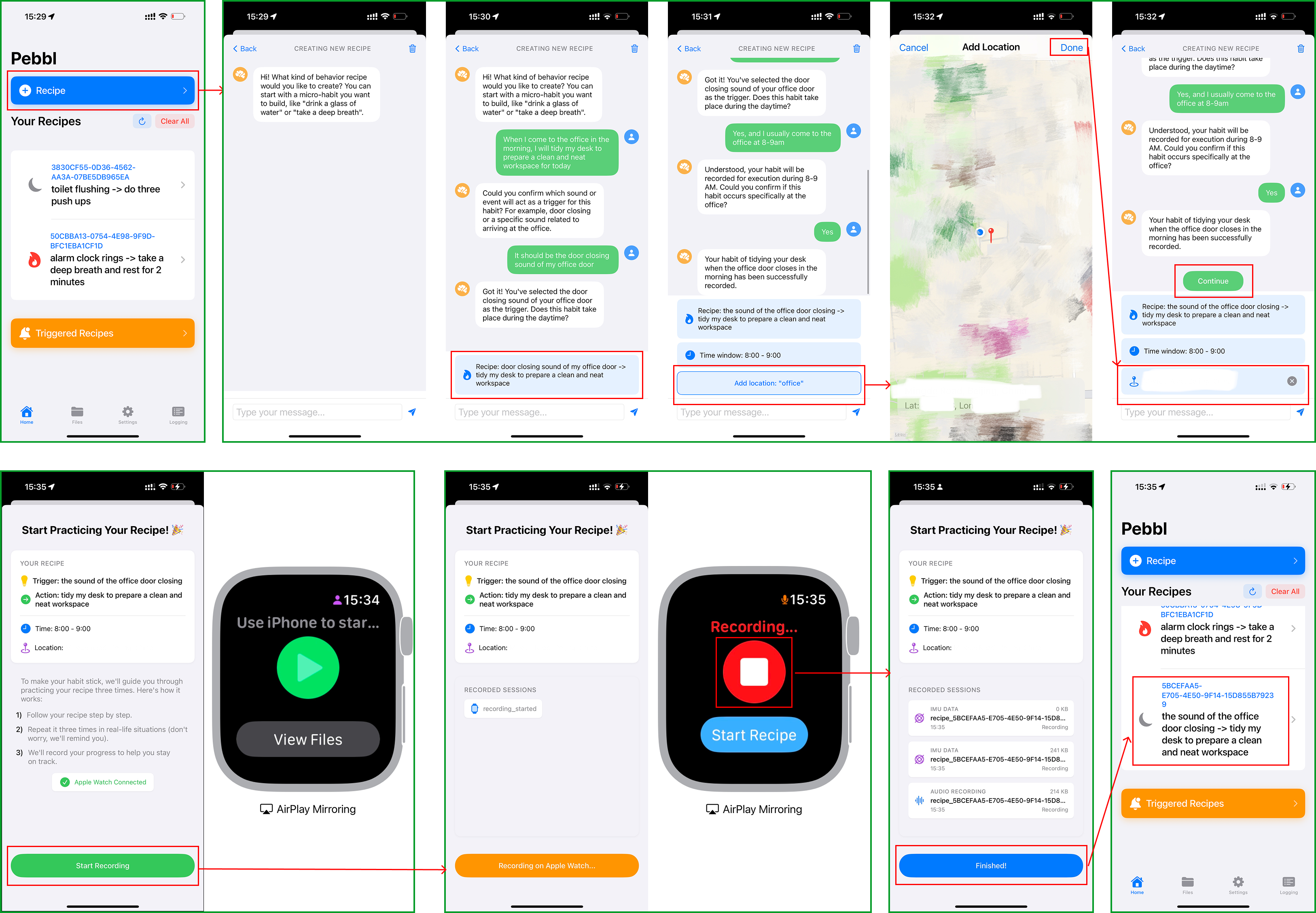}
    \caption{Detailed interaction flow and UI for creating recipes and practice.}
    \label{fig:append_detailUI_create}
\end{figure}

\autoref{fig:append_detailUI_create} shows the detailed UI and interaction flow when creating recipes and the optional practice process. Specifically, the home tab listed all recipes that the user has created (row1,col1). Users can click the \texttt{+Recipe} button on top of the home tab to navigate to the chatbot UI. Through the conversation process (row1,col2-4), the agent confirms the recipe, time condition, and location condition. When each required information is confirmed, \name automatically pops up a stacked list at the bottom. The location condition is a clickable button that navigates to the map UI (row1,col5) for the user to choose an accurate location with coordinates. After confirming the location, \name navigated back to the chatbot UI, and the location button turned into an information block. When all the required information is attained, \name pops up a \texttt{Continue} button for users to proceed to the practice stage (row1,col6).

In the practice UI(row2,col1-2), \name displays the detailed information overview on the top and requires the user to open the watch companion app. After clicking the \texttt{Start Recording} button on the phone, the watch app will begin recording (row2,col3-4). The user starts to practice the recipe several times and clicks the \texttt{Stop} button on the watch when the practice is done. Then, \name begins to transfer the recorded files to the phone and shows a \texttt{finished!} button (row2,col5) that navigates back to the home tab. The recipe list in the home tab shows the recently created recipe with a moon icon (row2,col6), indicating that this recipe is not active because conditions are not met.

Detailed UI and interaction flow for stages 3 and 4 (recipe execution and reflection) can be found in \autoref{fig:methods_user_journey} and \autoref{fig:itw_implement}. 

\subsection{Baseline Conditions.} \label{append:impl_baseline}

\begin{figure}
  \centering

  \subcaptionbox{Annotation UI for baseline1.\label{fig:append_anno_baseline1}}
    {\includegraphics[width=0.5\linewidth]{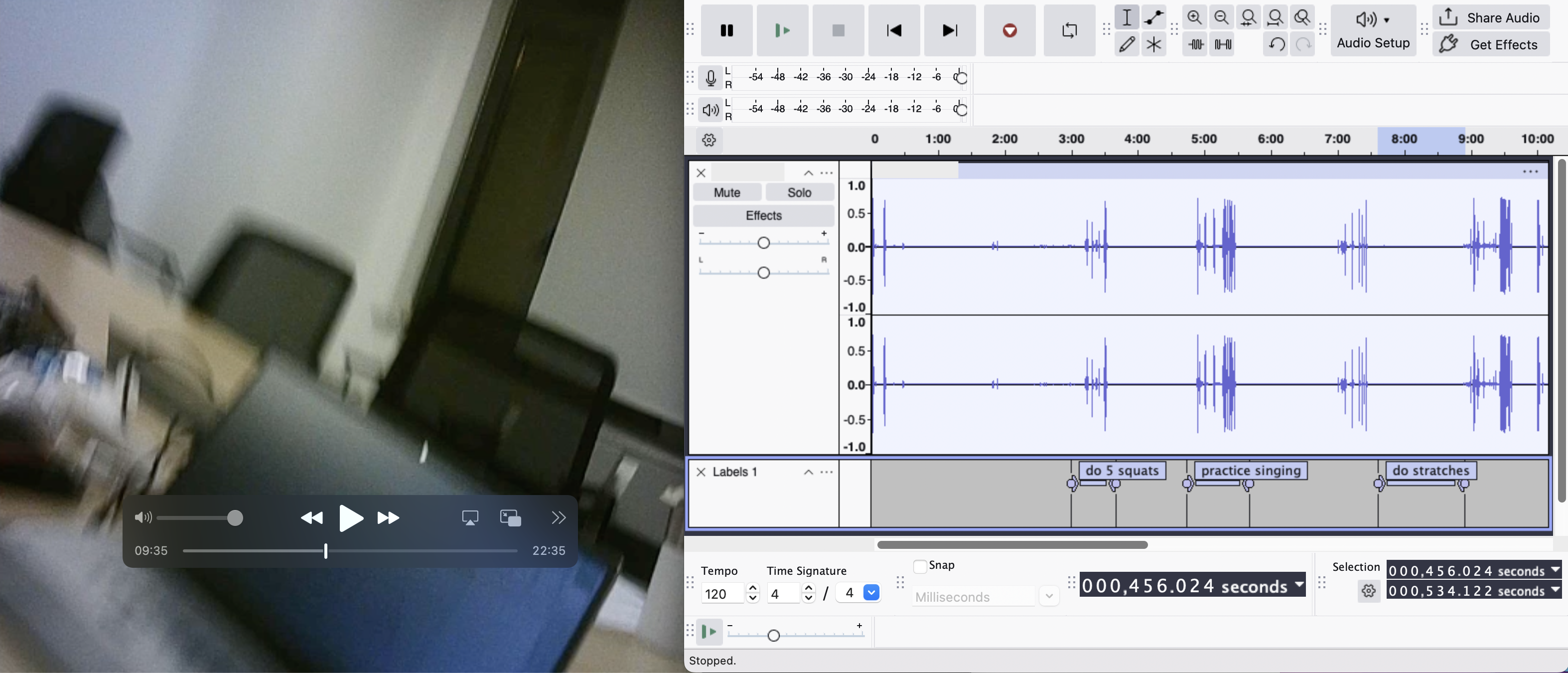}}
  \subcaptionbox{Annotation UI for baseline 2 and 3.\label{fig:append_anno_baseline23}}
    {\includegraphics[width=0.48\linewidth]{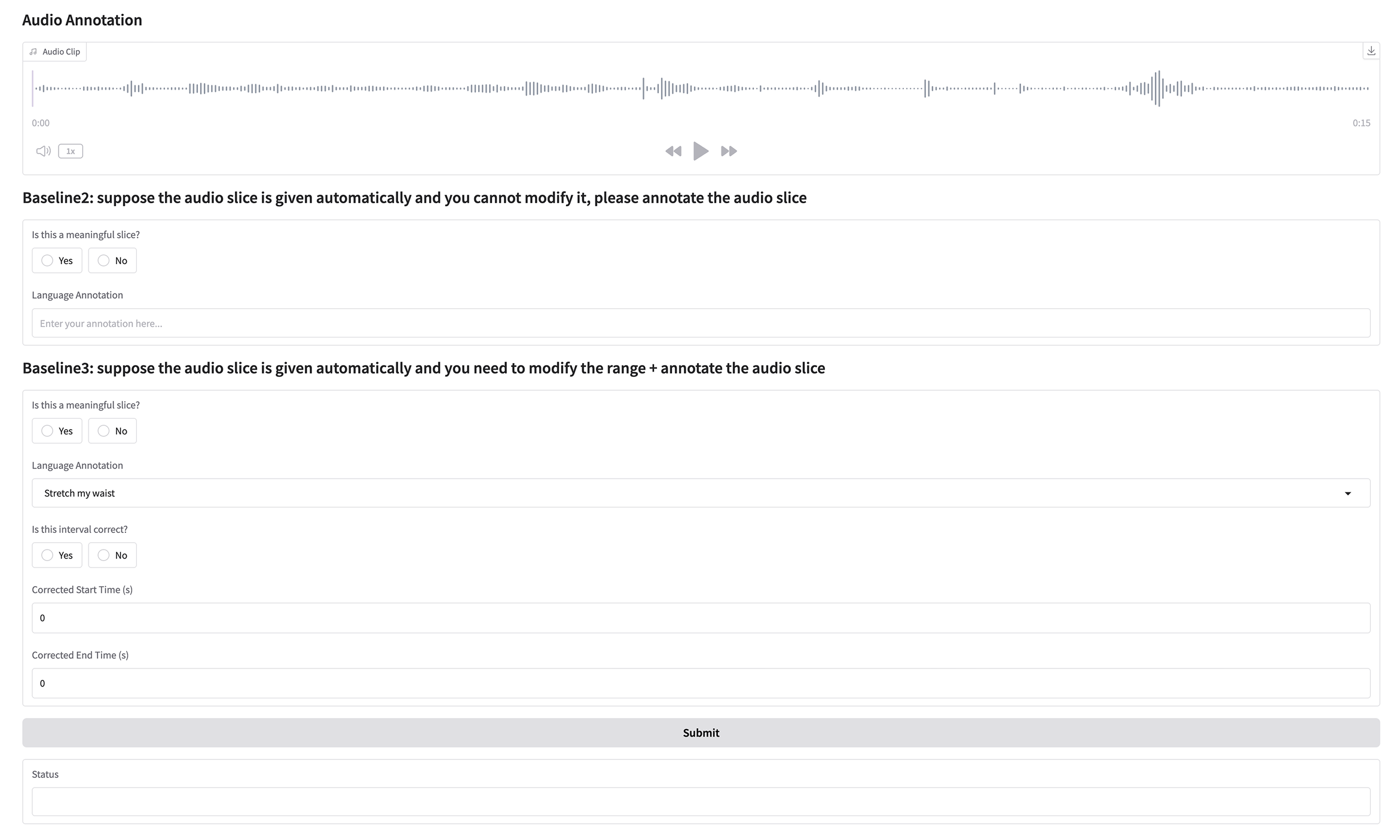}}
  
  \caption{Baseline conditions.}
  \label{fig:append_anno_baselines}
\end{figure}

Baseline conditions are as shown in ~\autoref{fig:append_anno_baselines}. We explain the baseline interaction protocol with these interactive UI and let participants try with one example. Specifically for baselines 2 and 3, we implement an illustrative webpage with Gradio.

\subsection{Trigger Choice Justification.} \label{append:impl_trigger_choice}

This section documents how we selected the four acoustic triggers used in the current implementation: door closing, toilet flushing, running water, and phone ringtone/alarm. \name's audio classification is built on Apple's SoundAnalysis framework, whose stock classifier (\texttt{SNClassifySoundRequest} with \texttt{classifierIdentifier: .version1}) recognizes over 300 everyday sound classes\footnote{Apple, ``Discover built-in sound classification in SoundAnalysis,'' WWDC 2021 Session~10036. \url{https://developer.apple.com/videos/play/wwdc2021/10036/}}. The four-class subset described here is a scoping decision for this paper's evaluation, not a property of the underlying pipeline; we did not train, fine-tune, or otherwise modify the audio classifier itself. The goal was to identify non-speech sounds that (1) are common across households, (2) are reliably detectable using the stock classifier, and (3) offer a clean temporal anchor for a wide range of tiny actions without requiring semantic understanding of speech or messages.

We began with the appendix of BJ Fogg's Tiny Habits, which lists roughly 300 example recipes~\cite{fogg2020tiny}. For each recipe, we judged whether the trigger is naturally associated with a distinct environmental sound. We excluded any triggers whose detection would require speech transcription or content understanding (e.g., ``when I get an invitation to a bar,'' ``when I says X'') to avoid severe privacy issue. For remaining candidates, we evaluated detectability against the stock classifier and action generality. Detectability refers to the presence of a distinctive acoustic signature with reasonable signal-to-noise ratio in everyday indoor settings. Action generality refers to the ability to precede many different actions to form an open-vocabulary activity dataset. Applying the above steps narrowed the set to four cues we used in current implementation.

Our current selection is a minimal, conservative set to establish feasibility. The underlying classifier already covers over 300 sound classes, expanding the trigger vocabulary requires deployment-time configuration paired with robust validation. The same pipeline could incorporate additional non-speech sound categories in future iterations (e.g., electric toothbrush motor for after brushing teeth, razor buzz for after shaving) without retraining the stock classifier, but each added class would still require context-specific robustness validation, threshold/gating tuning, and deployment testing.

\subsection{System Architecture and Trigger Detection Pipeline} \label{append:impl_pipeline}

\begin{figure}
    \centering
    \includegraphics[width=\linewidth]{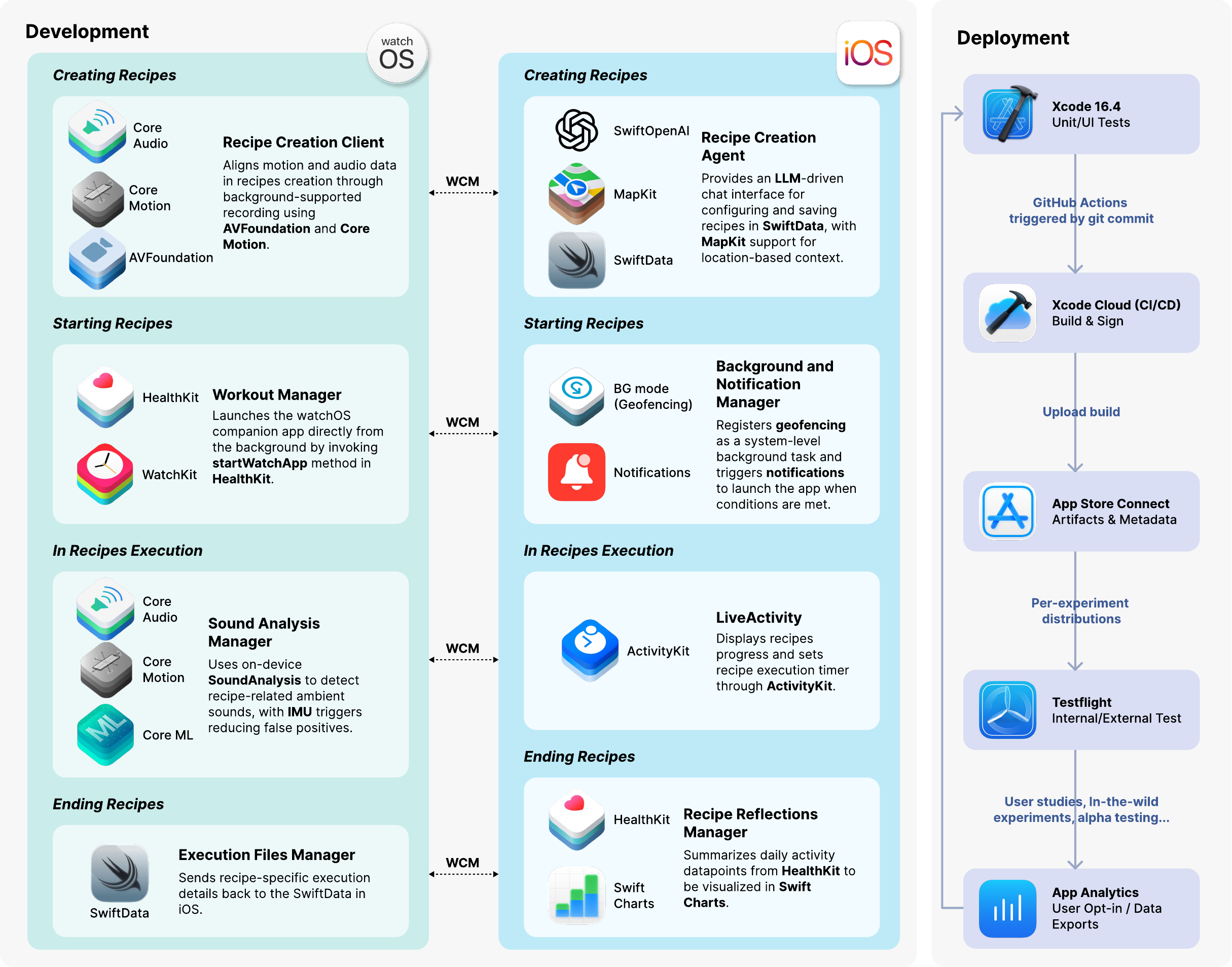}
    \caption{\name's system architecture and deployment workflow. The Watch Connectivity Manager (WCM) orchestrates data flow between iOS and watchOS.}
    \label{fig:methods_implementation}
\end{figure}

\begin{figure}
  \centering
  \subcaptionbox{A diagram showing the flow of data between endpoints.\label{fig:methods_data_flow}}
    {\includegraphics[width=0.65\linewidth]{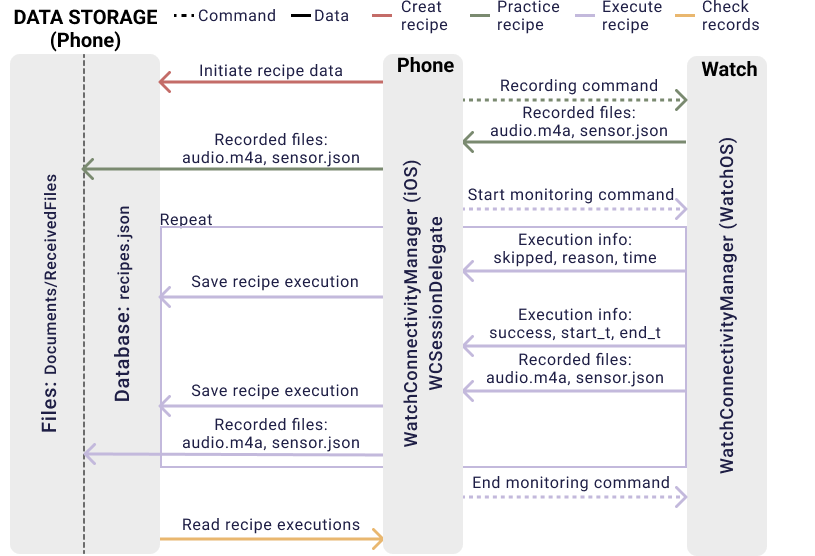}}
  \subcaptionbox{The database schema.\label{fig:methods_database}}
    {\includegraphics[width=0.3\linewidth]{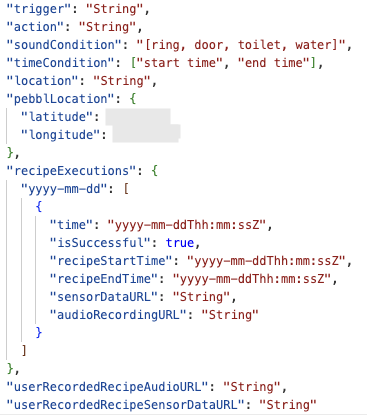}}
  \caption{\name's data management architecture across smartphone, smartwatch, and permanent storage.}
  \label{fig:methods_data}
\end{figure}

This appendix details the implementation of \name's cross-device applications, including framework choices, the system architecture (\autoref{fig:methods_implementation}), the data management pipeline (\autoref{fig:methods_data}), and the motion-gated audio trigger detection logic (\autoref{fig:in_recipes_execution_overview}).

\autoref{fig:methods_implementation} illustrates the development kits we leveraged to implement \name's core functionality, while \autoref{fig:methods_data} illustrates the flow of data across devices. The architecture prioritizes on-device processing, privacy preservation, and battery efficiency while maintaining real-time responsiveness for capturing in-the-wild behaviors. We decompose the system into four key phases of a recipe's lifecycle: creation, initiation, execution, and completion.

\subsubsection{Creating Recipes.}

Recipe creation is supported through a chatbot interface powered by Azure GPT-4o\footnote{\url{https://learn.microsoft.com/en-us/azure/ai-foundry/openai/concepts/models}} and the \texttt{SwiftOpenAI} package\footnote{\url{https://github.com/jamesrochabrun/SwiftOpenAI}}. Within this interface, users specify a recipe's trigger, action, location conditions, and time conditions through natural language dialogue with the agent. When a location condition is described, the agent provides a navigational button that directs the user to a \texttt{MapKit}-supported interface for precise location selection. Through this process, the agent extracts key parameters and encodes them into a structured recipe representation using \texttt{SwiftData}. The recipe metadata includes several fields: time conditions as ranges, auditory triggers as one of a few possible sound event categories, spatial conditions as GPS coordinates, and execution records as an initialized empty list.

\subsubsection{Starting Recipes.}

We employ multiple triggering mechanisms to ensure recipes activate at appropriate times and locations while minimizing battery consumption and false activations. The mechanism combines background monitoring, geofencing, and cross-device coordination.

\paragraph{\textbf{Geofencing and Background Monitoring.}}
Geofencing allows \name to remain dormant until relevant location contexts are detected. In our implementation, we use the \texttt{CLMonitor} to create geofences 200~m around recipe locations. When the device enters a monitored region, \name evaluates all location-enabled recipes to further determine if time conditions are met. Apple's \texttt{Background Tasks} framework further ensures \name can perform essential maintenance even when not actively running. The system schedules daily refresh tasks to prepare for upcoming recipe windows, update geofences, and schedule notifications.

\paragraph{\textbf{Cross-Device Recipe Activation.}}
We use \texttt{HealthKit} for recipe activation across devices. \name initiates data recording sessions using \texttt{HKHealthStore.startWatchApp()}, launching the watchOS app in a way that enables data recording without requiring users to interact with the watch's screen. When a session starts, \name begins on-watch sound detection and motion monitoring based on the active recipe configurations.

\subsubsection{Recipe Execution.}

\begin{figure}[ht]
    \centering
    \includegraphics[width=0.95\linewidth]{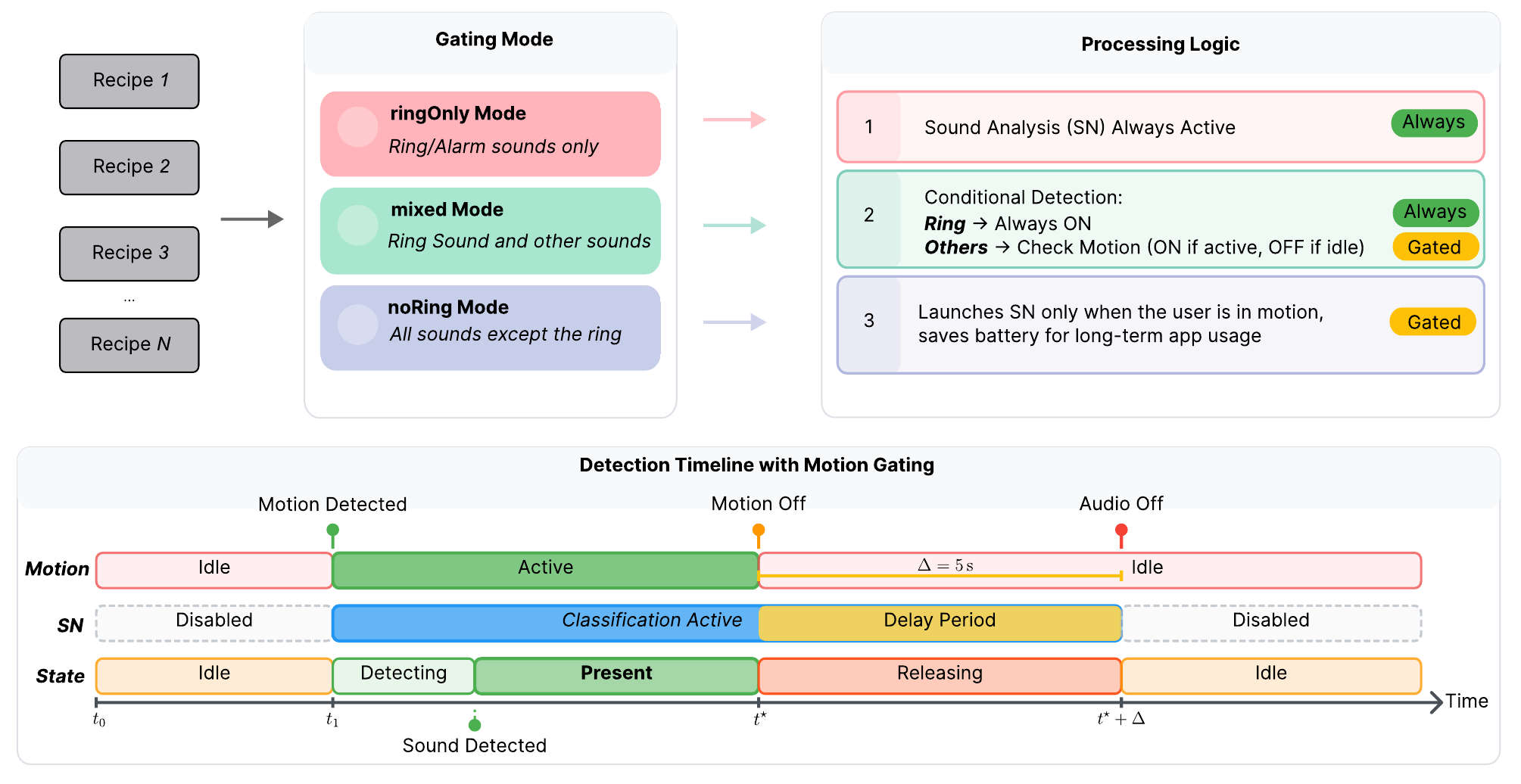}
    \caption{Motion-gated sound detection in \name's watchOS implementation. We assign each recipe's audio trigger with one of three gating modes: \texttt{ringOnly}, where sound detection remains continuously active; \texttt{mixed}, where ring sounds are always detected regardless of motion state, but other sounds are gated until the IMU indicates an \textit{active} state; and \texttt{noRing}, where all sounds are subject to motion gating.}
    \label{fig:in_recipes_execution_overview}
\end{figure}

During recipe execution, the system performs continuous real-time multi-modal sensing for trigger detection. Our implementation optimizes detection accuracy and battery efficiency through motion-gated audio classification and adaptive sensor activation strategies, as depicted in \autoref{fig:in_recipes_execution_overview}.

\paragraph{\textbf{Audio Pipeline.}}
We stream 16-kHz mono-channel audio into a classifier using Apple's \texttt{Sound Analysis} Framework\footnote{\url{https://developer.apple.com/documentation/soundanalysis}}. Each inference operates on a 1.5-second window with $90\%$ overlap and a real-time buffer of 4,096 data samples. The classifier outputs a confidence score $c_t \in [0,1]$ for each of the four sound events mentioned earlier: door closing, toilet flushing, running water, and ringtones or alarms.

\paragraph{\textbf{Motion Gating.}}
In our early development, we observed that sound classification can suffer from false positives due to poor subject awareness (e.g., people other than the user closing the door in the background). To mitigate this issue, we introduced a motion-gating strategy that not only suppresses false detections but also reduces battery consumption. Specifically, \name maintains a sliding window of $N=10$ accelerometer and gyroscope samples spread evenly across 2~seconds. At any moment in time $t$, we calculate the overall motion magnitude as $M_t = \big\|\mathbf{a}_t\big\|_2 + \omega \big\|\boldsymbol{\gamma}_t\big\|_2$, where $\mathbf{a}_t \in \mathbb{R}^3$ is acceleration, $\boldsymbol{\gamma}_t \in \mathbb{R}^3$ is angular velocity, and $\omega$ weights rotational motion. Motion at time $t$ is considered \textit{active} when $M_t$ exceeds a threshold $\tau_m$ for a minimum fraction $\rho$ of the entire window, as follows:
\begin{equation}
    \mathrm{Active}(t) \;=\; \mathbb{I}\!\left[\;\frac{1}{N}\sum_{i=0}^{N-1} \mathbb{I}\!\left(M_{t-i} \ge \tau_m\right) \;\ge\; \rho \;\right] \label{eq:motion-gating}
\end{equation}

\noindent
In our deployment, $\tau_m=0.5$ and $\rho=0.3$.

\paragraph{\textbf{Motion Gating Modes.}}
Since only some of the candidate audio triggers are associated with motion, \name implements three gating modes:
\begin{enumerate}
\item \texttt{ringOnly}: Motion gating is not used, so audio is always active,
\item \texttt{mixed}: Motion gating is used for all sounds other than ring sounds, and
\item \texttt{noRing}: All sounds are motion-gated.
\end{enumerate}
Whenever an audio trigger is motion-gated, audio classification activates once sufficient detection is motivated, and it deactivates 5~seconds after motion stops to accommodate brief pauses in activity.

\paragraph{\textbf{Detection State Machine.}}
To enhance detection robustness, we adopt a hysteretic approach with asymmetric thresholds. Specifically, let $c_{t,k} \in [0,1]$ be the system's confidence that the current sound belongs in class $k$ at time $t$. In order for an event to be detected, we require that the confidence score exceeds a threshold $\theta_p=0.5$ for a duration of $n_p = 2~\text{frames} \approx 300~\text{milliseconds}$. Meanwhile, we decide that an event is complete once the confidence score falls below $\theta_a=0.3$ for a duration of $n_a=30~\text{frames} \approx 4.5~\text{seconds}$. The state update for class $k$ is as follows:
\begin{equation}
S_t(k) \;=\;
\begin{cases}
\texttt{Present}, & (S_{t-1}(k)\!\in\!\{\texttt{Detecting},\texttt{Absent}\}) \wedge (n_p = \sum_{i=0}^{n_p-1}\!\mathbb{I}\!\left(c_{t-i}\! \ge\! \theta_p\right))\\
\texttt{Releasing}, & (S_{t-1}(k)=\texttt{Present}) \wedge (c_t \le \theta_a)\\
\texttt{Absent}, & (S_{t-1}(k)=\texttt{Releasing}) \wedge (n_a = \sum_{i=0}^{n_a-1}\!\mathbb{I}\!\left(c_{t-i}\! \le\! \theta_a\right))\\
\texttt{Detecting}, & (S_{t-1}(k)=\texttt{Absent}) \wedge (c_t > \theta_p)\\
S_{t-1}(k), & \text{otherwise}
\end{cases}
\label{eq:state-update}
\end{equation}
\noindent
Transitioning into \texttt{Present} while motion gating has been adequately satisfied initiates a trigger event.

\paragraph{\textbf{Multi-modal Data Recording.}}
During recipe execution, the sensor manager coordinates synchronized multi-modal data capture. Audio recording utilizes \texttt{AVAudioEngine} to record audio in the compressed AAC format at 48~kHz with a 64~kbps bitrate, balancing quality with storage efficiency. Accelerometer and gyroscope data are collected at 50~Hz. \name also maintains extended background runtime through \texttt{WKExtendedRuntimeSession}, automatically restarting sessions if they expire during recordings. All data processing occurs on background queues to prevent UI blocking, with thread-safe mechanisms for updating recording state and file management.

\subsubsection{Ending Executions.}

Whenever recipe executions end due to completion or manual stopping, the watch immediately stops recording and transfers files to the phone via the Watch Connectivity Manager (WCM). Metadata on completion status, event timestamps, the trigger, and optional location samples is applied to the execution record in the database.

\section{Experiment Detail}

This appendix collects per-study supporting materials: introductory materials shown to participants, example recipes authored during the studies, device details, the Study~2 post-study questionnaire, supplementary Study~2 analyses (the price-item results and the per-participant wrong-trigger breakdown), and the interview analysis method together with representative Study~3 quotes.

\subsection{Introductory Materials.} \label{append:exp_intro}

In this section, we show the introductory materials we used in Study 2 and 3. In practice, our experimenters guide participants with examples in this material.

\begin{tcolorbox}[enhanced, breakable,
    colback=gray!5!white,
    colframe=gray!50!black,
    title=Experiment Reference Material,
    fonttitle=\bfseries
]

\textbf{Theoretical Background} \\[0.5em]
The Fogg Behavior Model, proposed by Stanford psychologist B.J. Fogg, is a psychological framework for explaining human behavior and designing strategies for behavior change. This experiment builds on the Fogg Behavior Model to design a wearable-based habit formation app that helps users achieve their behavioral goals in a scientific and effective way.

Our research team applies the ``recipe for behavior change'' from the Fogg Behavior Model, guiding users to structure their target habit as a \textbf{``trigger + action''}. Target behaviors are simplified into small, easily achievable actions and anchored onto existing habits. This lowers the ability threshold and leverages familiar contexts as triggers, making behaviors more likely to occur.

\textit{Examples:}
\begin{itemize}
    \item After I pack my office bag, I will spend 5 minutes tidying my workspace.
    \item After I wash my lunchbox, I will put on my walking shoes.
    \item After I return to the office, I will get a cup of hot water.
\end{itemize}

\textbf{App introduction and recipe creation guidance:}

\textit{Approach 1 (for beginners):}
\begin{enumerate}
    \item What habits do I want to build? (e.g., exercise daily, stay organized, maintain a good mood)
    \item How can these habits be broken down into tiny-habits? (e.g., the starting action for running is putting on shoes; the simplified version of tidying is spending 3 minutes clearing the desk)
    \item Which trigger fits best with the tiny-habit? (consider location, frequency alignment, or the end of a triggering event, such as washing a lunchbox after eating lunch)
    \item When and where do I want to execute this recipe?
\end{enumerate}

\textit{Approach 2 (for experienced users):}
\begin{enumerate}
    \item What frequent activities in my daily life could serve as triggers?
    \item What positive habits could I anchor onto these triggers?
    \item When and where do I want to execute this recipe?
\end{enumerate}

\vspace{0.5em}
\textbf{Form fields:}
\begin{itemize}
    \item \textbf{Trigger:} Describe the triggering event rather than the sound.  
          Example: ``After I finish using the toilet and flush'' instead of ``toilet flush.''
    \item \textbf{Action:} Describe the micro-habit action.
    \item \textbf{Time:} The intended time window for executing the recipe.  
          Example: ``After washing my lunchbox after lunch (12–2 pm), I will put on walking shoes.''  
          Or: ``After flushing the toilet, I will do 3 push-ups (only between 1–3 pm).''  
    \item \textbf{Location:} The intended place for executing the recipe.  
          Example: ``After washing test tubes, I will tidy the reagent shelf'' (only in the lab).  
          Example: Push-ups after flushing the toilet are better at home than in the office.
\end{itemize}

Participants specify time and location according to their own routines.  
In the app, the location field retains the natural description (e.g., ``bathroom''), but the map view uses the laboratory address. The time description is recorded in the form, while the app entry uses the designated experimental time slot.

\end{tcolorbox}

\subsection{Recipe Examples Authored by Users.} \label{append:exp_recipe_examples}

In this section, we list the recipes designed by participants in Study2 and Study3, as shown in~\autoref{tab:append_exp_recipe_eg}. The actions shown in the table are limited by the objects that we can provide for in-lab experiments.

\begin{table*}[!t]
    \centering
    \caption{Recipe Examples Authored by Users.}
    \label{tab:append_exp_recipe_eg}
    \resizebox{\textwidth}{!}{%
    \begin{tabular}{|l|l|l|l|l|l|}
    \hline
    \textbf{Acoustic Cue} & \textbf{Trigger} & \textbf{Location Condition} & \textbf{Time Condition} & \textbf{Action} \\
    \hline

    \multirow{18}{*}{Ring} & When I hear the ring bell & in my offic & at night/afternoon & I will stretch my body for 2 minutes \\
    \cline{2-5}
     & When the alarm clock goes off & in my bedroom & in the morning (7.30am) & I will get off the bed and stretch my waist \\
    \cline{2-5}
     & When I hear the door bell & at home & anytime & I will look through the peephole \\
    \cline{2-5}
     & When the alarm clock goes off & in my bedroom & in the morning & I will get a cup of hot water \\
    \cline{2-5}
     & When the alarm clock goes off & in my dorm & in the morning (7-8am) & I will set a new alarm clock \\
    \cline{2-5}
     & When the alarm clock goes off & in my dorm & in the morning (8.30am) & I will pack my schoolbag \\
    \cline{2-5}
     & When the alarm clock goes off & in my bedroom & in the morning (6-10am) & I will get off the bed and do three squats \\
    \cline{2-5}
     & When the alarm clock for nap goes off & in lab & in the afternoon (2pm) & I will open the laptop and start working \\
    \cline{2-5}
     & When the alarm clock for activity goes off & in office & at daytime (10am-8pm) & I will get up and walk outside the door \\
    \cline{2-5}
     & When the alarm clock goes off & in my bedroom & in the morning (7-9am) & I will get up and open the curtains \\
    \cline{2-5}
     & When the alarm clock goes off & in my bedroom & in the morning (7am) & I will open the game to feed my chickens \\
    \cline{2-5}
     & When the alarm clock goes off & in my bedroom & in the morning (8.20am) & I will turn of the alarm immediately and get off the bed \\
    \cline{2-5}
     & When the alarm clock for bedtime goes off & in my dorm & at night (11-12pm) & I will un-plug my phone and laptop \\
    \cline{2-5}
     & When the timer on my desk goes off & in office & in the morning (9-12am) & I will check my work email \\
    \cline{2-5}
     & When the alarm clock for work goes off & at home & in the morning (8-9am) & I will leave home \\
    \cline{2-5}
     & When the alarm clock goes off & at home & in the morning (7.30am) & I will sit still for 3 minutes \\
    \cline{2-5}
     & When the alarm clock goes off & at home & in the morning (9.30am) & I will walk my dog \\
    \cline{2-5}
     & When I hear the ring for class over & at teaching building & in the morning (8-10am) & I will go running \\
    \hline
    \multirow{13}{*}{Water} & After I wash my hand & in office & at daytime (9am-5pm) & I will look far into the distance to relax my eyes \\
    \cline{2-5}
     & After I wash my hand & in office/dorm & at noon/evening & I will apply hand cream \\
    \cline{2-5}
     & After I wash my hand & in the canteen & in the morning (7-8am) & I will open Netflix \\
    \cline{2-5}
     & After I wash my hand & in office & at evening (5-7pm) & I will close my laptop \\
    \cline{2-5}
     & After I wash my cup & in lab & in the morning (10am) & I will get a cup of coffee/water \\
    \cline{2-5}
     & After I get a cup of water & in lab & at daytime & I will drink approximately 100ml \\
    \cline{2-5}
     & After I wash my hand in the washroom & at home & in the morning (7.30am) & I will wipe my hands with tissue \\
     \cline{2-5}
     & After I wash my hand & in lab & in the afternoon (4pm) & I will get a bag of snack from my backpack and eat \\
     \cline{2-5}
     & After I wash my cup & in office & in the afternoon (2-6pm) & I will walk around for 2 minutes \\
     \cline{2-5}
     & After I wash my hands before experiment & in lab & in the afternoon (2-3pm) & I will clean the experiment table \\
     \cline{2-5}
     & After I wash my face & at home & at night (11pm) & I will wipe my face using towel \\
     \cline{2-5}
     & After I wash my hand & in office & around noon (12-1pm) & I will water the plants \\
     \cline{2-5}
     & After I wash some fruits & in my dorm & around noon (11am-1pm) & I will share them with my friends \\
    \hline
    \multirow{4}{*}{Toilet} & After I flush the toilet & anywhere & at daytime & I will put down the toilet lid \\
    \cline{2-5}
     & After I go to the toilet & in office & in the afternoon (2-3pm) & I will relax my neck \\
    \cline{2-5}
     & After I flush the toilet & in office/at home & at daytime & I will use soap to clean my hand \\
    \cline{2-5}
     & After I go to the toilet & anywhere & at daytime & I will do heel lift exercise for 10 times \\
    
    \hline
    \multirow{19}{*}{Door} & When I come back to dorm & in my dorm & at night & I will do squats for 5 times \\
    \cline{2-5}
     & When I go back home from work & at home & at the evening (7pm) & I will read book \\
    \cline{2-5}
     & When I left my room & at home/in office & morning and night & I will check if I have locked the door \\
     \cline{2-5}
     & When I close the door & at home/in office & at night/morning & I will unpack my bag and clean up my desk \\
     \cline{2-5}
     & After I close the door (left the dorm) & at dorm & in the morning (9am) & I will check my bag to ensure my ID card \\
     \cline{2-5}
     & After I close the door & at dorm & at night (10-11pm) & I will go get my wash basket \\
     \cline{2-5}
     & After I close the door & at home & at night (7-10pm) & I will open window for ventilation \\
     \cline{2-5}
     & When I hear a door closing sound & at home & at daytime & I will check my mood and draw smiling face \\
     \cline{2-5}
     & When I go back to the dorm & at dorm & at night (8-12pm) & I will take out my laptop from backpack \\
     \cline{2-5}
     & After I go back to the dorm & at dorm & at night (9pm) & I will text my families \\
     \cline{2-5}
     & After I close the door & in my dorm & at night (9-11pm) & I will take out my key and phone from the pocket \\
     \cline{2-5}
     & After I close the door & in office & at noon (0.30pm) & I will take a nap \\
     \cline{2-5}
     & When I left the dorm & at dorm & in the morning (8-10am) & I will throw the trash \\
     \cline{2-5}
     & When I go home from work & at home & in the evening (7-10pm) & I will open TV \\
     \cline{2-5}
     & When I left for exercise & in my dorm & at evening (8-9pm) & I will close the air conditioner with a remote control app \\
     \cline{2-5}
     & After I close the door & in my bedroom & at night (11.30pm) & I will close the light and goto bed \\
     \cline{2-5}
     & After I come back home & at home & at evening (7pm) & I will wash my hand \\
     \cline{2-5}
     & After I left the office & in office & at evening (6-7pm) & I will go get some water \\
     \cline{2-5}
     & After I close the door & at home & at evening (5-7pm) & I will turn on the lights \\
    \hline
    \end{tabular}
    }
\end{table*}

\subsection{Device Details.}

We listed the device pairs' information in ~\autoref{tab:append_devices_pebbl}.

\begin{table}[h!]
\centering
\caption{Devices used during development and experiment.}
\label{tab:append_devices_pebbl}

\begin{tabular}{|l|l|llll|}
\hline
\textbf{Usage} & \textbf{ID} & \textbf{Phone Model} & \textbf{iOS Version} & \textbf{Watch Series} & \textbf{watchOS Version} \\ \hline
Development & D1 & iPhone 16 Pro Max & 18.6.2 & Series 10 & 11.6.1 \\
Development & D2 & iPhone 14 Pro & 18.5 & Ultra & 11.5 \\
 \hline
Study~2 & D3 & iPhone 13 Pro Max & 18.6.2 & Series 7 & 11.6.1 \\
Study~2 & D1 & iPhone 16 Pro Max & 18.6.2 & Series 10 & 11.6.1 \\
 \hline
Study~3 & P-D1 & iPhone 12 Pro Max & 18.5 & Series 10 & 11.6.1 \\
Study~3 & P-D2 & iPhone 16 Pro & 18.6.2 & Series 8 & 11.6 \\
Study~3 & P-D3 & iPhone 13 Pro Max & 18.6.2 & Series 7 & 11.6.1 \\
Study~3 & P-D4 & iPhone 13 & 18.5 & SE 2 & 11.6 \\
Study~3 & P-D5 & iPhone 16 & 18.6.2 & Series 7 & 11.6.1 \\
Study~3 & P-D6 & iPhone 14 & 18.6.2 & Series 9 & 11.6.1 \\
Study~3 & P-D7 & iPhone 12 Pro & 18.5 & Series 9 & 11.6 \\
Study~3 & P-D8 & iPhone 16 Pro & 18.6.2 & Series 10 & 11.6.1 \\
 \hline

\end{tabular}
\end{table}

\subsection{Study 2 Post-Study Questionnaire} \label{append:study2_questionnaire}

Mental- and physical-demand items were adapted from NASA-TLX; the remaining items were authored by the research team to probe adoption and annotation constructs specific to \name's two comparison classes. All items used a 7-point Likert scale unless otherwise noted.

\textbf{Reminder/tracking-side items (administered for the comparison between \name and the commercial tracking app, B4).}
\begin{enumerate}
    \item \textbf{Willingness to use:} How likely are you to use this habit-tracking approach?
    \item \textbf{Mental demand:} How mentally demanding is it to use this habit-tracking approach? \textit{(adapted from NASA-TLX)}
    \item \textbf{Physical demand:} How physically demanding is it to use this habit-tracking approach? \textit{(adapted from NASA-TLX)}
    \item \textbf{Recommend to friends:} How likely are you to recommend this habit-tracking approach to your friends?
    \item \textbf{Stick to use:} How likely are you to stick to using this habit-tracking approach (i.e., how likely are you to grow a habit of using this app)?
    \item \textbf{Usefulness for habit:} How useful is checking tracked records for fostering your habits? For \name, we additionally asked participants to score the usefulness of checking \emph{skipped} execution records.
\end{enumerate}

\textbf{Annotation-side items (administered for the comparison between \name and the three annotation baselines B1--B3).}
\begin{enumerate}
    \item \textbf{Mental demand:} How mentally demanding is it to use this data collection + annotation approach? \textit{(adapted from NASA-TLX)}
    \item \textbf{Physical demand:} How physically demanding is it to use this collection + annotation approach? \textit{(adapted from NASA-TLX)}
    \item \textbf{Prefer to use:} How do you prefer this collection + annotation approach?
    \item \textbf{Label confidence:} How confident are the labels you produced through this annotation approach?
    \item \textbf{Price:} What is the price that you would like to sell at for 100 pieces of data? \textit{This item was rated without a scale restriction and normalized later for analysis. We explicitly required participants to imagine themselves as users of \name\ who had already produced 100 data records through usage. Results and construct-validity discussion for this item are reported in Appendix~\ref{append:study2_price}.}
\end{enumerate}

\subsection{Study~2: Expected-Compensation (``Price'') Item: Results and Construct Validity} \label{append:study2_price}

\textbf{What the item measured and how it was administered.} The exact wording is reproduced verbatim in Appendix~\ref{append:study2_questionnaire} (annotation-side item~5). Participants answered with a free numeric response rather than a Likert rating, imagining themselves as users of \name\ or of a baseline workflow (B1--B3) who had just produced 100 data records, and they rated each condition after that condition's exposure (the full \name\ session and the brief trials of B1--B3 described in Section~\ref{sec:study2}, Post-experiment). Before answering, participants were told that existing crowdsensing approaches compensate contributors for labeled data, so that they could respond with that framing in mind; we did not, however, anchor them to a specific platform or rate. Raw responses were normalized per participant by dividing each condition's value by the minimum non-zero value that participant reported across the four conditions, so that the cheapest workflow chosen by each participant equals $1.0$ and the others are expressed as multiples of that minimum. This normalization is necessary because, in the absence of a locked external anchor, participants naturally chose their own units and scales, and a per-participant normalization is the mildest intervention that makes the four conditions comparable within each participant.

\textbf{Aggregate results.} After normalization, participants set the lowest price for \name\ ($1.16$) relative to B1/B2/B3 ($3.16$/$2.18$/$2.05$; Wilcoxon signed-rank, $p=9.73{\times}10^{-5}$, $1.13{\times}10^{-2}$, $8.33{\times}10^{-4}$). The gap was largest against B1 (post-hoc video annotation) and smaller but still present against B2 and B3. The direction is consistent with the self-reported mental and physical demand findings for the same annotation-workflow comparison (Section~\ref{sec:study2}, annotator-side subjective results): conditions rated as more effortful were also priced higher. Interview comments indicate that participants did not treat price as decoupled from effort; for example, P8 and P14 explicitly described combining workload and quality when setting a number.

\textbf{What we do and do not conclude.} We interpret these numbers as a \emph{perception} signal about how costly each workflow feels to a participant who has just performed it, and no more than that. In particular, we do not treat them as a market-calibrated wage, as evidence about the downstream value of the labels produced, or as a recommendation for how crowdsensing platforms should compensate contributors. Two construct-validity concerns warrant this scoping. First, although participants were told that crowdsensing approaches compensate contributors, they were not anchored to a specific platform rate; values therefore still reflect each participant's own estimate of fair compensation under limited exposure, which is also why the analysis had to normalize per-participant rather than compare raw values directly. Second, \name\ received substantially more exposure in Study~2 than any baseline, and additional contact with a workflow could plausibly raise or lower its perceived cost through mechanisms this single item cannot disentangle (fatigue, practice, anchoring). For both reasons we report the price finding here as a supporting observation that is \emph{directionally consistent} with the effort measures, rather than as an independent claim, and the main paper's narrative does not lean on it.

\subsection{Study~2: Per-Participant Wrong-Trigger Distribution} \label{append:wrong_trigger_per_participant}

\begin{figure}
    \centering
    \includegraphics[width=0.8\linewidth]{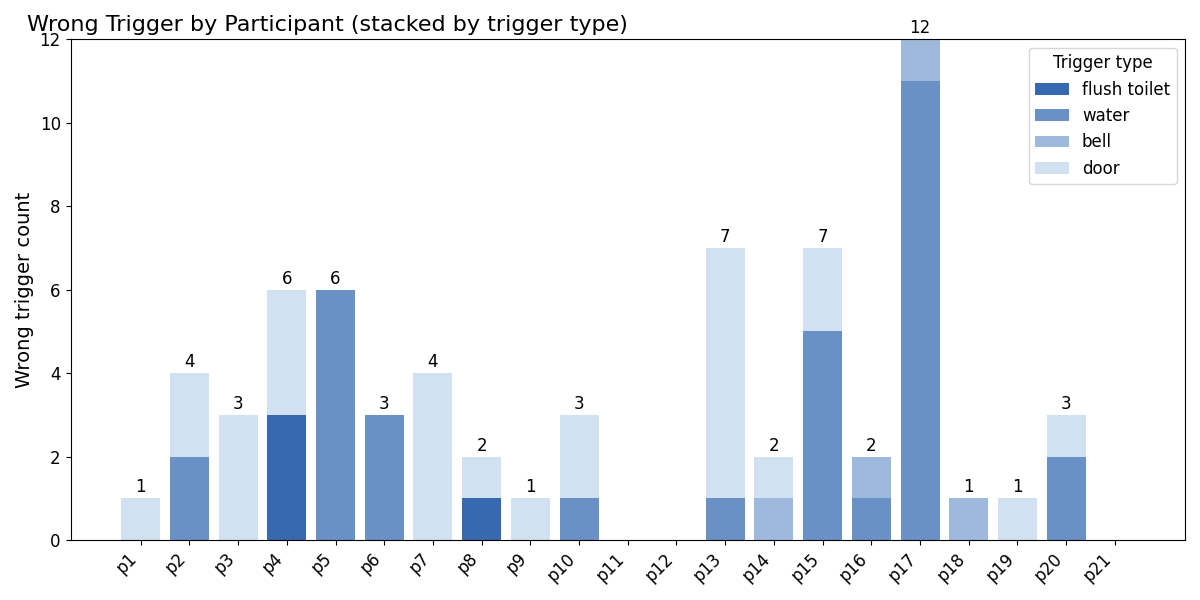}
    \caption{Per-participant wrong-trigger counts, stacked by trigger class. Same 68 events as the per-category view in main-paper \autoref{fig:eval_wrong_trigger_count}.}
    \label{fig:append_wrong_trigger_participant}
\end{figure}

\autoref{fig:append_wrong_trigger_participant} reports the same 68 wrong-trigger events summarized in Section~\ref{sec:exp2_wrong_trigger_analysis}, broken down by participant and stacked by trigger class. Errors are dispersed across individuals: most participants show $\le 3$ wrong triggers, while a small tail accounts for the bulk (e.g., P17=12; P13/15=7), and the stacks within those outliers are dominated by \textit{water} and \textit{door} --- the same two noisy classes identified in the main text. This distribution is consistent with the lab environment in Section~\ref{sec:study2}: experiments were conducted within $<10$\,m of a washroom, kitchen, elevator room, and the building's waste area, so sessions during busier daytime windows encountered more domestic-water and door-closure confounders than sessions in quieter windows.

\subsection{Interview Analysis.}

We describe the analysis method we used for both Study~1 and Study~3 interviews, followed by representative Study~3 quotes.

\subsubsection{Interview Analysis method.} \label{append:exp_interview_braun}
For both Study1 and Study3, we roughly followed Braun and Clarke’s six-phase reflexive thematic analysis~\cite{braun2023doing} to make sense of the interview data. Two authors lead the interviews and conducted the analysis. After data collection, we read the full transcripts and highlighted segments that met one of two criteria: (1) the same or highly similar point was raised by multiple participants; or (b) the statement was insightful or informative for understanding \name and its use context. For each highlighted quote, we attached a short topical label capturing the core idea in the participant’s own terms and then clustered related labels. 

When iterating the thematic structure for Study1, we organized expert's core idea under 4 predetermined themes - pros, cons, concerns, future refinements. For Study3, we intentionally aggregated the positive feedback about \name into a single theme, while allocating more granular themes to concerns, limitations, and forward-looking expectations. We made this choice because the paper already details \name’s advantages through design exposition and quantitative results; in contrast, readers are likely to gain more from a differentiated view of limitations and opportunities for improvement emerging from the interviews.

\subsubsection{Study3 Interview Quote.} \label{append:exp_interview_quote}

\textbf{Users like \name for its intelligence, sense of feedback, ease of use, and usefulness for habits.} Overall, participants valued \name's ``smart'' automation (P8) and tight watch–phone integration(P4/5), which made capturing moments feel effortless and timely.

\begin{quote} 
    P8: \textit{``It (baseline) can't automatically remind you; you have to open it manually, swipe, and log things yourself. It needs proactivity (from the user), which I am not. It is not intelligent, but yours is, in an automated way. With your method, it feels more like having an AI assistant, a bit like ‘Hey Siri.’ I can just double-tap my fingers, and it starts recording right away, super convenient. The whole thing gives me a sense of intelligence.''}

    P4: \textit{``One big advantage is, it links the watch and the phone, that’s very convenient. I had to bring my phone when I'm running to complete a record, but normally I only take my watch when I run. With your app, that's a lot more practical.''}
    
    P5: \textit{``Say, in the morning, I’m washing my hands, I don't wanna pick up the phone to log something, because my hands are wet... Logging on the phone is already a hard habit to form... your method, the interaction is simple, I can use it even with wet hands.''}
\end{quote}

The haptic feedback created a rewarding sense of interaction (P3).

\begin{quote}
    P3: \textit{``a sense of interaction... When the system detects something automatically and gives a vibration or a ‘ding,’ it is a rewarding kind of feedback.''}
\end{quote}

Several participants (P6/7/8) noted that \name's gentle ``pressure'' to act was actually productive, kick-starting action and sustaining habit formation in everyday contexts. 

\begin{quote}
    P6: \textit{``Because of this reminder, I feel like I really need to do it, but the baseline app just doesn't matter. But this will cause, with your method, (I am) more likely to act, and that helps the habit stick. It’s not bad (the pressure). It (baseline) cannot provide supervision because there is no pressure. With your methods, it indeed continuously reminds me to finish this habit, and it is useful, not dust-laden in my app list.''}

    P7: \textit{``It’s really easy to forget... Like drinking water, some people just don't like it...That's why Apple Watch have stand reminders for long sitting, sometimes you sit, feel comfortable, or get absorbed in your work, just completely forget. Some small but important behaviors in daily life might need it more (\name), like behavioral habits, not big (habits) like living style, it's different.''}

    P8: \textit{``Why I think baseline is low in physical demand, because if I forget (record), it feels like all the practice was for nothing, then I will lack motivation to act (do the habit). But with yours, the watch alerts me, then I know I need to do it. It is like forcing me into action, it's a physical burden. I think burden is food, it has a positive effect instead... I think, after all, forming a habit needs to start doing it.''}
\end{quote}

\textbf{Audio-cue based detection is suitable for behavioral habits.} 
Participants (P4/7) framed audio-triggered detection as a strong fit for small, repeated, context-bound behaviors, such as drinking water, brief exercise sets, and breaking up sedentary time.

\begin{quote} 
    P4: \textit{``I also use it, baseline is suitable for once-a-day habits, like I do sports today, or practicing calligraphy. Yours fits better for small things, like five squats or push-ups at a time, scattered. I might use both. That’s why I gave similar scores for both... Because most of my habits are once a day, it just needs to be recorded... Yours has its own benefits; if I need repeated ones, I can set it hourly, as the start or the end of the day, which also has its scenarios.''}

    P7: same quote as the previous paragraph.
\end{quote}

On the other hand, once-a-day ``macro'' habits were not suitable for automatic detection due to a higher wrong trigger rate (P1/2). Therefore, in order to capture broader activity events, \name needs to add support for macro habits as well, sacrificing label granularity to a certain degree. 

\begin{quote} 
    P1: \textit{``I found that your app needs sounds, water, ringtone. If it is running in the background, it (wrong triggers) might suddenly pop up in a meeting.''}

    P2: \textit{``It (\name) is not very accurate, sometimes, I need to dismiss multiple times (the fans)... It (baseline) is not suitable for fragmented habits, but like running 30 minutes, manual record works fine. But like fragmented habits, by the end of the day, you just forget how many times you did it.''}
\end{quote}

\textbf{\name has a steep learning curve and users need guidance to design recipes.} 
\name has a steep learning curve because users need to understand what tiny-habit is, how to use \name, and the initial setup felt demanding (P3/5). Also, users need design guidance when they first start to use \name (P2/3/7): they were unsure what habits to target, which cues to choose, and how to translate vague intentions into workable recipes. Users asked for stronger scaffolding, such as a conversational agent that can co-design recipes in real time by turning fuzzy ideas into concrete, triggerable routines. This introduces a challenge: how can we appropriately design this agent that can turn this need into a chance of increasing vocabulary openness?

\begin{quote}

    P3: \textit{``Baseline feels like requiring you to make a plan up front, and planning is the hardest part for someone like me... Your method is a bit better, but it still needs me to set up a recipe in advance. It would be better to have recommendations, let's say, if I have a vague idea, you could help me make it concrete and turn it into a recipe... Yes, the chat agent should be more powerful, like chat with me in real time, and help resolve this while we talk. Also, sometimes I don’t really know what I want, and I’m not sure what kinds of behaviors the software can record or trigger under what conditions, so it would be better if the agent could give clearer recommendations.''}

    P5: \textit{``Because when I start training, the operations feel a bit complicated, connecting the watch to the phone, something... The first time I used it, it was a little difficult, but overall it was interesting and felt like a good way to start forming habits.''}

    P2: \textit{``Right now I haven’t decided what kind of habit I want to build, so I can’t think of a sound trigger I’d like to use... Honestly, when I first got the app, I was pretty confused about what habit to set, what recipe to make.''}

    P7: \textit{``For the watch, when you set the pipeline at the beginning, you need to think, what small thing can I use as a reminder for a bigger habit? Later on, once it reminds you, you just record it.''}
\end{quote}

\textbf{Failure records can assist reflection, yet might lead to frustration.}
Participants appreciated comprehensive logs and the motivational lift of seeing successful entries (P2/7), but several cautioned that surfacing failures can dampen morale (P6). P7 regards the usefulness of seeing failure records as culturally related.

\begin{quote}
    P2: \textit{``Seeing successful records makes me happy; seeing failure records can feel unpleasant...''}

    P7: \textit{``For habit formation, both (positive and negative feedback) are useful. If you found out you are good at something, you get positive feedback, and you reinforce it. Like in fitness, when you see results, you are more likely to continue. Negative feedback can sometimes feel like criticism; it might make you want to perform better. I think it (whether negative feedback is motivating might) also depends on cultural differences.''}

     P6: \textit{``I think if I used this software for a long time, if I had many failures early on but then improved later, that could be motivating. But if I saw too many failures at the beginning, I might don't want to (continue). Still, it's a bit more useful compared to showing only successes.''}
\end{quote}

They suggested balancing visibility with supportive design: aggregate views over time, emphasize trends and recoveries, and pair outcomes with a collection mechanism (P1/2).

\begin{quote}
    P1: \textit{``Like, give some badges? Yes, yes, I like collecting badges, so I’d be very willing. Just like with XXX Coffee (a brand), I always check in, if there’s a pop-up event, I’ll go buy a cup.''}

    P2: \textit{``Maybe you could do something like the app Forest, rewards for success and penalties for failure... I think I prefer persisting for several days and getting a reward. If you persist longer, the rewards should become more valuable.''}

\end{quote}

\textbf{Users hope \name could add support for more triggers.}
Participants wanted richer and more personalized triggers beyond the current audio set, ideally allowing user-defined cues (P2), broader sensing (e.g., sedentary time, timers), and context awareness to reduce mistimed prompts (P5). They also suggested triggers for health routines (e.g., medication, P4), screen-time control, and social/accountability features (P5/7). They also asked for smarter feedback summaries beyond visualization to explain why a trigger fired and how to adjust behavior (P7).

\begin{quote}
    P2: \textit{``Your app has a lot of improvement space, right now it only recognizes three sounds... I'm thinking of recording my own triggers, I can provide a few samples, also some description, so that you can recognize more sounds... I can't think of a trigger sound I want, but if you only have water, door, and ringtone, it feels limited... The alarm clock, can the watch detect, for example, I’ve been sitting for 40 minutes and remind me to move.''}

    P4: \textit{``I need to take a kind of pill in the long term, every day at fixed times. It’s not exactly habit-building, it’s just a real need... Adding this would be very practical, especially for older users, to help them manage chronic conditions.''}

    P5: \textit{``Sometimes the trigger fires, but it’s not the right moment to do the habit. It will be better if it can detect whether it is the correct timing, combining contextual information, it would feel more convincing. But if it is not, I might feel like my current context is not for it (the habit). For example, I have two places to wash my hands at home... I only want to use (the trigger) at the bathroom. But it also reminds me at the kitchen.''}

    P5: \textit{`·At work, we often sit for long periods, some habits like standing, or get some water to drink would be useful, health-related. Apple Watch has something similar; it sends a notification to remind you to stand, but it’s easy to ignore. I don't know, maybe for me, my motivation is not enough, it just tells me to stand, but it doesn't have enough feedback mechanism... I’d prefer (feedback) that’s visible on my phone, not just a small watch screen. Maybe not only charts or numbers as a summary, ideally, AI could summarize this data and tell me, such as how long I have sat and what can be the consequences, in a very straightforward way, might draw my attention... Numbers and charts, maybe interesting for the first time, but that can become meaningless over time, I can get numb to it.''}

    P5: \textit{``Also screen time. Sometimes I get absorbed in short videos or browsing, time just flies... Strong intervention is disrupting because I might really be doing something; if it is just a reminder, I might just skip it. It would be better if the app could redirect my attention away from the screen to something else, so I can actually take a break.''}

    P5: \textit{``Exercise, mine is relatively regular, I might go every other day... Because I have colleagues going with me, I think when someone else is with you, it might be more effective. If you put this into it (\name), you can execute (recipes) with your friends, this might increase the success rate? ''}

    P7: \textit{``I think for small habits, current triggers are enough. But for others, like exercise, a timely notification each day might be sufficient... Still, automatic pipelines, like playing with my phone, I need to set an alarm when I begin to play, then it can trigger the reminder. If it could recognize when I pick it up and automatically set a reminder, which might work better.''}
\end{quote}

\end{document}